\documentclass[natbib]{svjour3}                     
\smartqed  
\usepackage{graphicx}
 \usepackage{aps-bibstyle}  
\newcommand {\apgt} {\ {\raise-.5ex\hbox{$\buildrel>\over\sim$}}\ }
\newcommand {\aplt} {\ {\raise-.5ex\hbox{$\buildrel<\over\sim$}}\ } 
\begin{document}

\title{Solar surface and atmospheric dynamics
}
\subtitle{The Photosphere}


\author{Mart\'\i nez Pillet, V.      
}


\institute{Instituto de Astrof\'\i sica de Canarias \at
              38200, La Laguna, Tenerife, Spain\\
              Tel.: +34-922605237\\
              Fax: +34-922605210\\
              \email{vmp@iac.es}           
}

\date{Received: date / Accepted: date}

\maketitle

\begin{abstract}
Various aspects of the magnetism of the quiet sun are reviewed. The suggestion
that a small scale dynamo acting at granular scales generates what we call the
quiet sun fields is studied in some detail. Although dynamo action has been
proved numerically, it is argued that current simulations are still far from
achieving the complexity that might be present on the Sun. We based this
statement not so much on the low magnetic Reynolds numbers used in the
simulations but, above all, in the smallness of the  kinetic Reynolds numbers employed by
them. It is argued that the low magnetic Prandtl number at the solar surface
may pose unexpected problems for the identification of the observed
internetwork fields with dynamo action at granular scales. Some form of
turbulent dynamo at bigger (and deeper) scales is favored. The comparison
between the internetwork fields observed by Hinode and the magnetism inferred from
Hanle measurements are converging towards a similar description. They are both
described as randomly oriented, largely transverse fields in the several
hecto-Gauss range. These similarities are ever making more natural to assume
that they are the same. However, and because of the large voids of magnetic
flux observed in the spatial distribution of the internetwork fields, it is
argued that they are not likely to be generated by dynamo action in the
intergranular lanes. It is concluded that if a dynamo is acting at granular
scales, the end product might have not been observed yet at current spatial
resolutions and sensitivities with the Zeeman effect. Thus an effort to
increase these resolutions and polarimetric sensitivities must be made. New
ground- and space-based telescopes are needed. The opportunity offered by the
Solar Orbiter mission to observe the Quiet Sun dynamics at the poles is seen as
one of the most important tests for confirming the existence, or otherwise, of
a granularly driven surface dynamo.  
\end{abstract}

\section{Introduction}
\label{sec:1}

A consensus about the existence of a small-scale dynamo (SSD\footnote{Small
scale here refers to generation of magnetic fields at scales smaller than the
energy injection one, the granulation.}) operating at the solar photosphere is
being consolidated in today's solar physics \citep[see, e.g.,][for a
review]{Vogler2007,Abbett2007, Pieta2010, Stein2012}.  High resolution
magnetograms from ground and (mostly) space-based telescopes observed in the
internetwork are often used to indicate that such a surface dynamo exist
\citep[e.g.][]{Danilovic2010a, Lites2011}.  On theoretical grounds, simulations
of various kinds have been used to suggest how universal the various
ingredients of such a dynamo seems to be \citep{Moll2011}. They all indicate
that turbulent shear stresses acting on the inertial range act as the main
mechanism able to efficiently convert kinetic into magnetic energy. The
conclusions from the various simulations of turbulence in a conducting fluid
seems to be that it would have been a lot harder to explain the absence of a 
surface SSD than its presence.  With an emphasis on the observational side, 
we review in this work the current status of this consensus and try to pinpoint 
which aspects are more solidly established and which are less settled.

Section \ref{sec:2} will be the only one that concentrates on simulations and
the theoretical aspects related to the problem of the existence of SSDs on the
Sun. It will address the most controversial argument questioning the existence
of such a mechanism, namely the fact that the solar convective zone has a
magnetic Prandtl number that is orders of magnitude smaller than one, while
simulations work near the $\sim 1$ regime most of the time.  Low magnetic
Prandtl numbers are known to severely discourage dynamo action since the early
simulations of convectively driven turbulent dynamos \citep{Nordlund1992,
Scheko2004b,Scheko2005}. Recently, progress has been achieved, however, that
indicates that an SSD is indeed possible in the low magnetic Prandtl regime
\citep{Iska2007,Scheko2007,Branden2011}. But the situation is not conclusive
and the papers addressing this issue often resort to the fact that a mixed
polarity field is observed at the solar surface as the firmest indication that
such a mechanism should exists. However, and in the absence of a clear proof
that this observed (internetwork) fields originate from an SSD --and such a
proof is not available yet-- the only progress to settle this issue will
come from further work in the simulation front. 

The observational arguments that have been put forward to favor the presence
of a solar surface dynamo are discussed later in Section \ref{sec:3}.  There
are basically two such arguments. First, the evidence from the observed
Hanle effect in lines such as the SrI 4607 \AA~line and the careful modeling of
these signals indicate the existence of a tangled field with a mean strength
of $<B>\sim$ 130 G \citep{Trujillo2004} some few hundred kilometers above the
solar surface \citep[see also][and references therein]{Trujillo2006}.  This
number was originally derived under some model assumptions that made it
uncertain to within a factor two. However, a recent study
\citep{Shchu2011} of the predicted Hanle signals from the MHD simulations
described in \cite{Pieta2009a} has eliminated some of this model dependency and
confirmed such large mean field strengths.  As a mixed polarity tangled field
at unresolved scales leaves basically no trace in the Zeeman profiles, these
fields have always been a prime candidate to be considered as originated from a
surface SSD.  If the tangling occurs at scales near or above present
resolutions, some signatures can be detected, though.  It is unclear if these
fields have been observed using the Zeeman effect \citep[but see][and Section
\ref{sec:3}]{Lites2009, Bellot2012}.  The second observational argument in
favor of an SSD comes from the Hinode spectropolarimeter (SP) instrument
\citep[][]{Tsuneta2008, Kosugi2007} and its unprecedented characterization of
the internetwork fields using the Zeeman effect \citep[][]{Lites2008,
Ishikawa2011, Orozco2012}.  While transverse fields were known to exist in the
quiet sun, as originally found by the Advanced Stokes Polarimeter
\citep[ASP;][]{Lites1996}, it was totally unexpected that these fields have a
predominant transverse character. This transverse nature seems to fit in a
natural way with an origin related to a turbulent dynamo 
as shown by recent simulations \citep[see][]{Schussler2008}. 

A possible outcome given this state of affairs could be as follows. The
existence of an SSD acting at granular scales at the solar surface can
eventually be confirmed from a set of improved SSDs simulations (along the
lines described in Section \ref{sec:2}). In them, a continuous distribution of
fields is obtained that is able to explain the Hanle depolarization levels from
those fields created at the smaller scales and the largely horizontal
internetwork fields from those at larger scales.  The separation between these
two sets of fields does not have to be sharp and a range of spatial scales can
contribute to both the Zeeman and Hanle results \citep[or, perhaps, that the two observed
processes are due to fields exactly at the same scales as it can be inferred
from the recent results of][]{Bellot2012}.  Note that, such a field
distribution would solely depend on the existence of the always present
turbulent convective motions near the surface and, thus, should be independent
of latitude and of activity cycle phase. While this conclusion seems rather
plausible given the current evidence, the aim of this work is to address some
of the known problems that might prevent such an outcome. In particular, Section
\ref{sec:4} describes some observations recently obtained with the IMaX/SUNRISE
magnetograph \citep[][]{Pillet2011, Solanki2010} that show regions that display
very little magnetic activity, either measured as residual signals in
time-averaged deep magnetograms or as evidenced by a lack of flux emergence
episodes in the form of small-scale loops \citep[as discovered
by][]{Marian2012}. It is unclear how these voids are compatible
with a granularly driven SSD. That the situation is far from clear has been corroborated
recently by the study of \citep[][based on SDO/HMI magnetograms]{Stenflo2012}
who proposes the existence of a basal flux of order 3 G that is suggested to be
an upper limit to the efficiency of an SSD at the solar surface. According to
this result all the internetwork fields observed with Hinode/SP will not be
generated through such a mechanism and only the Hanle depolarizing fields 
could be originated through it (if at all).

In spite of this somewhat confusing situation, it is important to stress that our
understanding about the nature and the properties of the quiet sun fields has
improved enormously in recent years. But it is clear that a number of important
questions remains on the theoretical/modelling side and on the observational
front.  Section \ref{sec:5} finishes this work proposing a way forward to
further advance in this understanding of the quiet sun magnetism. Not
surprisingly, we promote an effort to increase the polarimetric sensitivity and
the spatial and temporal resolutions of both, the Hanle and the Zeeman
observations. Studying the statistics of the quiet sun fields at various
latitudes will also prove crucial.

\section{Small scale dynamo action at low $P_m$. Implications for the solar case}
\label{sec:2}

The seminal reference that triggered the present debate on the existence of a
convectively driven turbulent dynamo at the solar surface was the work of
\cite{Cattaneo1999}, although the debate is older \citep[see,
e.g.,][]{Petrovay1993, Lin1995}.  It is important to point out that this work
mentioned, both, the granular and supergranular scales as possible contributors
to such non-helical dynamo.  The simulations used closed upper and lower
boundaries with vertical fields in both of them. The Reynolds and magnetic
Reynolds numbers could be clearly defined for this simulation thanks to the
fixed computational grid used to solve the MHD equations. They were ${\mathrm
Re}=u l/\nu=200$ and ${\mathrm Re}_m=u l/\eta=1000$, with $l$ the
characteristic length of the energy injecting convective cells, $u$ the
velocity of these cells  and $\nu$ and $\eta$ the molecular viscosity and the
magnetic diffusivity, respectively. These numbers are large enough to ensure
the development of turbulence. But it is important to note that ${\mathrm Re}$ was
five times smaller than ${\mathrm Re}_m$ (magnetic Prandtl number of ${\mathrm
P}_m={\mathrm Re}_m/{\mathrm Re}=\nu/\eta=5>1$). Under these circumstances, the
magnetic field sees a smooth mean flow efficiently acting on it.  The numerical simulation
resulted in dynamo action saturating at 20\% of the kinetic energy flow.  The
crucial ingredient was the chaotic nature of the driving flows. Figure 2 of
this paper already showed that, at the surface, the strong fields were
localized in the downflow lanes, while the cell interiors showed no (vertical)
field signature. The situation was different in deeper layers where fluctuating
fields were filling basically the whole volume. At the time of the
publication, the dominant transverse nature of the internetwork was not known
(see Section \ref{sec:3}) and this aspect was not analyzed. For this reason, the
profiles synthesized by \cite{Almeida2003} using these simulations concentrated
on the study of the asymmetries induced in the Stokes V profiles (circular
polarization) and its comparison with those observed in the internetwork. A
shortage of asymmetries indicated that the simulations still did
not achieve as much complexity as present in the Sun.  However, using this
synthesis, and after including effects such as telescope diffraction, it was
predicted that when improving the spatial resolution from 1 arcsec to 0.15
arcsec, one should detect four times more Stokes V polarization signals.

A number of assumptions made in this simulation (such as the Boussinesq
approximation) have been relaxed in more recent works. The more realistic ones
(in terms of their proximity to the physical conditions on the Sun) are those
made with the MURaM code \citep[][]{Vogler2007, Schussler2008, Pieta2010}.  In
particular, they have addressed the important question of the role played by
the closed boundary conditions assumed in \cite{Cattaneo1999}.
\cite{Stein2003} pointed out that, in the solar convective zone, fields are
submerged efficiently to the bottom of the convective zone by strong and
concentrated downflows showing little recirculation near the surface. This
recirculation was artificially enhanced in the simulations of
\cite{Cattaneo1999} by the use of closed boundary conditions. \cite{Stein2003}
concluded that diverging upflows sweep the fluid into downflows, often
vortical, where stretching  and twisting becomes effective (and balanced by
diffusion) but all these fields are rather rapidly submerged down into the bulk
of the convective zone. The energy added to the flux that visits the surface
was a very small fraction of the global budget of magnetic energy and the
effect cannot be considered a local dynamo. This criticism has, however, been
superseded by the MUraM simulations which used an open boundary and allow for a
non-zero pointing flux at the bottom boundary.  The way in which this boundary
condition was implemented in the simulations of \cite{Vogler2007} was by
imposing an artificially increased magnetic diffusivity there. This diffusivity
ensured that horizontal fields moving downward in the simulation leave the box
unimpeded while, at the same time, prevented horizontal flux from entering the
domain.  All flux leaving the bottom boundary in these simulations was created
by dynamo action inside the box. \cite{Vogler2007} concluded that while the
downward pumping of flux outside of the domain does indeed reduce the growth
rate of the dynamo, it does not shut it down.  They predict that as long as a
sufficiently high ${\mathrm Re}_m$ is used (above the critical magnetic
Reynolds number, ${\mathrm Re}_m^C$), dynamo action will be observed in the
simulations. Specifically, they see exponential growth of the magnetic energy
for ${\mathrm Re}_m$ of 2600, while they find a decrease for ${\mathrm Re}_m$
below 1300. Now, here we should caution that the exact values of ${\mathrm
Re}_m$ and ${\mathrm Re}$ (and thus of ${\mathrm P}_m$) are not as easily
defined as in \cite{Cattaneo1999} simulations. The need to resolve shocks with
artificial viscosity schemes or the implementation of the bottom boundary
diffusivity necessarily implies that these numbers are more difficult to
ascertain. An estimate of the effective Reynolds and Prandtl numbers for the
MURaM simulations was given by \cite{Pieta2010} using various moments of the
velocity and magnetic spectra (or Taylor microscales). The various dynamo runs
available from this code turn out to have ${\mathrm Re}_m\in [2100,8300]$ (with
the latter value using a grid resolution of 4 km) and ${\mathrm P}_m\in
[0.8,2]$.  This set of simulations all displayed an SSD generating magnetic
fields inside their volume. The conclusion was that current MURaM simulations
(with ${\mathrm P}_m\sim 1$) will show dynamo action as long as ${\mathrm Re}_m > 
{\mathrm Re}_m^C\sim 2000$.  The main mechanism identified for this
generation was the stretching and twisting of field lines by fluid motions in
the inertial range of the spectrum of velocity fluctuations.  In particular,
it was concluded that dynamo action is concentrated in the turbulent
downflows as field line stretching against magnetic tension is very efficient
there \citep[e.g.][for a review]{Stein2012}. In order to clarify the nature
of the observed dynamo action, \cite{Moll2011} have analyzed the underlying
physical mechanism under various physical conditions and assumptions. They
concluded that for the cases studied (incompressible MHD, Boussinesq
convection and compressible solar convection), the field is amplified by
similar inertial range shear stresses that are independent of the conditions
at the injection scale. The inclusion of compressibility effects or the
asymmetry between upflows and downflows generated by the strong
stratification did not influence the result. They, thus, termed the dynamo
mechanism as universal.

One concern remains about the existence of a possible SSD on the Sun, though.
The problem has been known since the early studies of dynamo action in
conducting fluids. It was originally formulated by \cite{Batchelor1950} who
studied how turbulent motions stretch the field lines and amplify the
magnetic energy as long as this process remains unimpeded by ohmic diffusion.
Field line stretching is produced by fluid motions in the inertial range whose
dissipation scale is set by the viscosity of the fluid. High viscosity $\nu$
couples the field lines to the flow and allows it to bend them efficiently.
Large magnetic diffusivity $\eta$ decouples the plasma (and the flows) from the
field lines and prevents the bending. Thus, it was always clear that the
efficiency of the SSD was going to be controlled by the interplay of these two
effects as measured by the ratio ${\mathrm P}_m=\nu/\eta$. Clearly, both
conditions ${\mathrm Re}_m \gg 1$ (to favor field line stretching) and
${\mathrm P}_m \gg 1$ (to couple fluid motions and field lines) boost local
dynamo action.  Based on an analogy between vorticity and magnetic fields
\cite{Batchelor1950} even concluded that for $P_m<1$ no SSD was possible. This
conclusion was later criticized by a number of authors as the analogy cannot
include the different initial and boundary conditions seen by these two fields
\citep[see, e.g.][]{Boldryrev2004}. However, simulations in the early 90's
\citep[see, e.g.,][]{Nordlund1992}, including compressibility and strong
stratification, already resulted in dynamo action only for ${\mathrm P}_m\geq
1$ with efficient shutting down of the dynamo for ${\mathrm P}_m< 1$. Other
simulations encountering the same problem are discussed in \cite{Boldryrev2004}
and in \cite{Scheko2005}.  Thus, the question of the existence of an SSD at low
Prandtl numbers has received some attention in recent years.

\begin{figure*}
  \includegraphics[width=1.\textwidth]{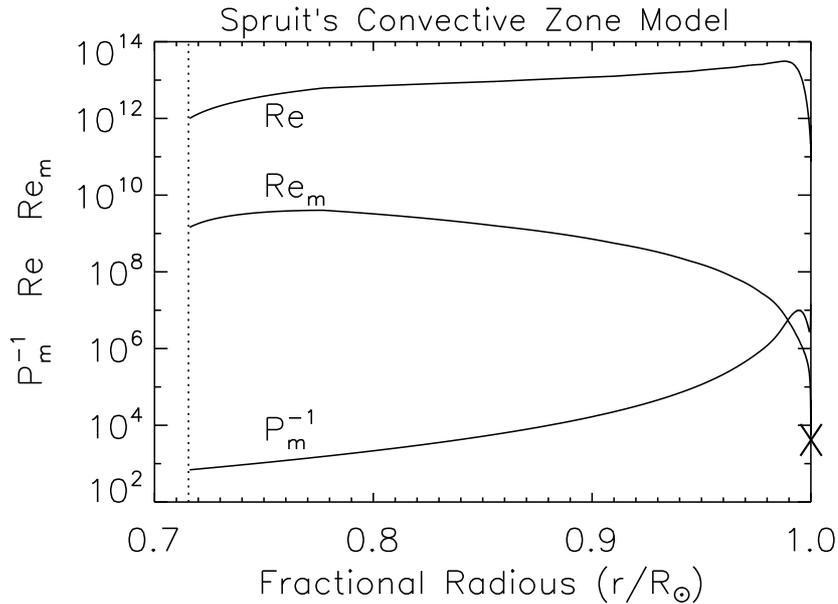}
\caption{The Reynolds number, magnetic Reynolds number and (inverse) Prandtl number
in the solar convective zone according to the mixing-length 
model of \cite{Spruit1974}. The X near the surface marks the tipical values achieved 
of both ${\mathrm Re}$ and ${\mathrm Re}_m$ in the simulations.}
\label{fig:1}       
\end{figure*}

Before we briefly describe the results from the numerical studies about the
existence of an SSD at low Prandtl numbers, it is important to remember what
are the actual numbers that occur in the solar convective zone and get an idea
of how far or how close are we from simulating these conditions. To this end,
Figure \ref{fig:1} shows ${\mathrm Re}$, ${\mathrm Re}_m$ and ${\mathrm
P}_m^{-1}$ as computed in the mixing-length based model of \cite{Spruit1974}.
The velocities at the injection scale $u$ are obtained as part of the model and
the characteristic length is assumed here to be $l=z/2$, with $z$ the depth
inside the convective zone. The magnetic Reynolds number changes from $10^5$ at
the photosphere to $10^9$ at the bottom of the convective zone, whereas the
Reynolds number stays constant at a level of $10^{12}$. This makes ${\mathrm
P}_m\in [10^{-7},10^{-3}]$, with the smallest value reached at the photosphere.
Thus, everywhere in the Sun, we have ${\mathrm Re}\gg {\mathrm Re}_m \gg 1$ and
${\mathrm P}_m\ll 1$ which is exactly the regime where the existence of an SSD
becomes problematic. It is generally believed that for a sufficiently high
$Re_m$ there will always be an SSD at work. But the above mentioned
simulations prompted a deeper study about the existence and nature of an SSD
under solar conditions.  As already mentioned, the simulations by
\cite{Cattaneo1999} had ${\mathrm P}_m=5$ and those from the MURaM code always
move close to the ${\mathrm P}_m\sim 1$ case (there is one with ${\mathrm
P}_m=0.8$ that is discussed below).  Note that the regime where ${\mathrm
Re}_m$ and ${\mathrm Re}$ are similar is actually very favorable for the
numerical codes as similar grid sizes resolve the dissipative scales of both,
magnetic and velocity fields.

What is the physical argument behind this difficulty to generate an SSD when
${\mathrm P}_m\ll 1$?. Under such conditions the viscous ($l_\nu$) and
resistive ($l_\eta$) scales follow $l_\eta/l_\nu \sim {\mathrm P}_m^{-3/4}\gg
1$ \citep{Scheko2004b, Scheko2005} and the resistive scale $l_\eta$ falls in
the middle of the inertial range.  Turbulent eddies of scales $l>l_\eta$ do the
necessary field line bending and twisting for dynamo action at a rate of
$u_l/l$ (with $u_l$ the typical flow velocity at this scale). If ${\mathrm
P}_m>1$ only these eddies occur and the field lines always see a spatially
smooth flow. However, if ${\mathrm P}_m<1$ one has now eddies below the
resistive scale $l<l_\eta$. These eddies act on the field as a turbulent
diffusion with diffusivity $u_l l$ and destroy magnetic energy. It is the
predominance of this last process what can make dynamo action impossible 
as it was seen in the previously mentioned simulations. 
Using an incompressible spectral MHD code and the PENCIL code
\footnote{See http://www.nordita.dk/software/pencil-code.}, \cite{Scheko2005} studied
what are the possible asymptotic limits when ${\mathrm Re}\gg {\mathrm Re}_m$
and the corresponding values of the ${\mathrm Re}_m^C$ for the existence of a
dynamo. The above described effect always translates into a sharp increase in
${\mathrm Re}_m^C$ as ${\mathrm P}_m \rightarrow 0$ \cite[see
also][]{Pieta2009b}, but does not prevent the existence of a dynamo in this regime.
The two asymptotic limits are, first, as ${\mathrm Re} \rightarrow \infty$,
${\mathrm Re}_m^C \rightarrow$ const, so that dynamo action is possible for
higher ${\mathrm Re}_m$ and, second, ${\mathrm Re}_m^C \rightarrow\infty$ with
${\mathrm Re}_m^C/{\mathrm Re}\rightarrow {\mathrm P}_m^C=$const, in which case
no dynamo is possible (turbulent diffusion efficiently dissipates magnetic
energy at small scales). Which exactly of the two asymptotic limits prevails
has been under much debate in recent years. While \cite{Scheko2004b} and
\cite{Scheko2005} (see their Figure 2) favor the existence of a ${\mathrm
P}_m^C$ (no dynamo) from simulations of incompressible magnetoconvection
reaching values of ${\mathrm P}_m$ as small as 0.15, \cite{Boldryrev2004}
provided analytical arguments favoring the existence of a ${\mathrm Re}_m^C$. 

While the debate in the mid last decade did not look promising for confirming
the existence of an SSD at the solar surface, the situation has changed
recently (even if not completely settled). \cite{Scheko2007} \citep[see
also][]{Iska2007} performed simulations of incompressible MHD turbulence
reaching values of ${\mathrm P}_m=0.1$ and with a sufficiently high ${\mathrm
Re}_m$ that indicated a plateau region where a ${\mathrm Re}_m^C$ is observed
(see their Figure 1b). Admittedly, the number of such simulations proving the
existence of this plateau is very small but the authors consider it enough
numerical certainty. The results from \cite{Branden2011} using the PENCIL 
code
(that includes compressible effects) and low values of ${\mathrm P}_m$ resulted
in dynamo action being activated as well.  \cite{Pieta2010} also find growth of
magnetic energy in the one case they analyzed with ${\mathrm P}_m=0.8$.  Thus,
the most advanced existing numerical simulations of small-scale dynamo action
in turbulent MHD currently favor the occurrence of such a process in the low
${\mathrm P}_m$ regime.  However, a number of caveats remain: 

\begin{itemize}
\item First, and most importantly, ${\mathrm Re}_m^C$ increases with decreasing
${\mathrm P}_m$. The exact factor depends on the specificities of the
simulations. \cite{Boldryrev2004} suggest a factor 7 increase in ${\mathrm
Re}_m^C$ when shifting from the ${\mathrm P}_m \gg 1$ to the ${\mathrm P}_m \ll
1$ case.  They use an analytical model of isotropic and homogeneous turbulence
that includes the extra roughness of the velocity field for ${\mathrm P}_m <
1$.  \cite{Scheko2007} find from incompressible forced turbulence a factor 3
increase. As in the MURaM simulations one has ${\mathrm
Re}_m^C\sim 2000$ \citep{Pieta2010}, this means that as soon as we move into 
the low-${\mathrm P}_m$ regime, we need magnetic Reynolds numbers above at least 6000 
to be able to trigger dynamo action. These magnetic Reynolds numbers are not currently 
achieved by this code. In particular, none of the runs used by \cite{Danilovic2010a}
would be able to actually sustain dynamo action.  On top of that, the
saturated field strength is known to decrease with decreasing ${\mathrm P}_m$,
thus the expected field strengths will be smaller than those computed for
${\mathrm P}_m \sim 1$.  The exact amount of this reduction is still a very
controversial issue \citep[see][]{Scheko2004b, Scheko2007, Branden2011} and its
magnitude for the solar case unknown. But the net effect will be a reduction in
the fields as compared to those computed for ${\mathrm P}_m > 1$.

\item To complete the demonstration of the existence of a dynamo driven by
fluid motions in the inertial range at low ${\mathrm P}_m$ values, a growth
rate of the dynamo scaling with ${\mathrm Re}_m^{1/2}$ must be obtained from
the simulations. Neither \cite{Scheko2007} nor \cite{Pieta2010} have reached
that (see Figure 3 of the latter work).  Simulations with an increased
resolution are needed to finally settle this issue.  \cite{Scheko2007}
concludes that, as long as this is not achieved, the mechanism that sustains
the growth of the magnetic field fluctuations in the low-${\mathrm P}_m$ regime
will remain basically {\em unknown}.

\item \cite{Scheko2007} and \cite{Iska2007} concluded that in the ${\mathrm
P}_m\ll 1$ regime, the magnetic energy spectra is fundamentally different from
that found in the ${\mathrm P}_m\gg 1$. The spatial distribution of the growing
magnetic fields is qualitatively different too \cite[see Figure 2
in][]{Scheko2007}. This indicates that the use of simulations in the ${\mathrm
P}_m\sim 1$ range to compute the ensuing Stokes profiles and its comparison
with those observed may not be justified.

\item The often simulated case with ${\mathrm P}_m\sim 1$ has the same
spectral energy properties and field distribution as the ${\mathrm P}_m\gg 1$
case \cite[][]{Scheko2004a}.  This is probably why the case with ${\mathrm
P}_m=0.8$ simulated by \cite{Pieta2010} was so similar to the those in the
range of ${\mathrm P}_m\in [1,2]$.

\end{itemize}

Let us finalize this section by stressing that if the situation looks
confusing, it is because this has indeed been the case in this topic for some
time \citep[see][who speak about a frustrating outcome]{Iska2007, Scheko2007}.
One argument commonly given to promote the existence of an SSD is the
observation of a mixed polarity field in the internetwork regions of the Sun
(and many of the above mentioned works use this argument one way or another).
Figure 2 of, both, \cite{Cattaneo1999} and \cite{Vogler2007} clearly indicate
that this is a reasonable argument. The point we want to stress here is that
the same applies to Figure 10 from \cite{Stein2006}, which {\em does not}
include an SSD.  \cite{Pieta2009b} estimate for this simulation a ${\mathrm
Re}_m\sim 600$, which is known to be too small to develop dynamo action. However, their mixed
polarity distribution located in the interganular lines looks as `solar' as in
the other cases. In the work of \cite{Stein2006} emphasis is made on diverging
upflows bringing flux to the surface, expulsion to the intergranular lanes and
sweeping of field lines into strong downflows that carry the flux into deeper
layers. These simulations extent typically further down than those that concentrate
in SSD generation and also include larger scales such as those associated with the
mesogranulation.  As shown in Fig. \ref{fig:1}, the deeper we move into the
Sun, the larger ${\mathrm Re}_m$ and ${\mathrm P}_m$ (although still smaller
than one).  Thus, a valid question is if it is not more natural to ask if a
solar SSD exists at meso- and supergranular scales and, if so, whether they
dominate over that that might exists at granular ones. This point will be
further discussed in Section \ref{sec:4}. 

\section{Observed signatures of small scale dynamo action at the solar surface}
\label{sec:3}

We now turn to the observational aspect of the discussion and ask the question:
Have we seen the fields produced by a possible SSD operating at the solar
surface?  As it will become evident, there has been, as before, solid observational
progress and areas with much confusion.  Basically, two candidates exist that
are often considered as by-products of an SSD, the internetwork fields and the,
so-called, hidden fields that generate the Hanle depolarization signatures.

\subsection{Zeeman signals}

There is no question that Hinode/SP data has produced a major quantitative and
qualitative jump forward in our understanding of the internetwork fields. The
publication by \cite{Lites2008} of a gigantic slit scanned map with consistent
$10^{-3}$ polarimetric sensitivity and homogeneous spatial resolution of 0.3
arcsec changed our view of the quiet sun magnetism. In this map, a myriad of
patches with predominant linear polarization signatures was discovered. The ASP
already found the existence of episodic burst of largely transverse fields
\citep[the Horizontal Internetwork Features, HIF,][]{Lites1996} but they were
thought to be rather sporadic.  The only previous indication of their existence
and global character came from the SOLIS instrument as found by
\cite{Harvey2007}. But no prediction from the simulations or estimate of their
magnitude was available.  The existence of this ubiquitous horizontal field
has now received full confirmation from the SUNRISE/IMaX data
\citep{Danilovic2010b} who could make the first study of their evolution
\citep[see the animation in][]{Solanki2010} and establish a solid statistics of
their lifetimes. Both instruments, Hinode/SP and SUNRISE/IMaX coincide in
locating these HIF at the borders of the upflowing granules for a large
fraction of their evolution. \cite{Lites2008} emphasized that the linear
polarization signatures were not co-spatial with line-of-sight fields that 
were more frequently found in the intergranular lanes.

Before the publication of \cite{Lites2008} most of the discussion on the
internetwork flux concentrated on obtaining its mean unsigned flux
\citep[see][for a review on this topic previous to the impact of the Hinode
measurements]{Solanki2009}. The idea was that an intricately complex field
with mixed polarities observed in the best available Stokes V magnetograms 
was the outstanding description of the internetwork fields. Increased
resolution (or sensitivity) will result into ever increasing amounts of
longitudinal signals observed as there was less and less cancellation due to
instrumental effects. Either because of lack of reliable measurements of the
field inclination or because of a habit to focus studies of solar magnetism
exclusively in longitudinal magnetograms, no mention to its possible transverse
character was traditionally made. As the field strengths were expected to be near or below
equipartition values \citep[less than 500 G, see][]{Keller1994}, these fields
were not thought to be necessarily vertical either. Before the Hinode results,
measurements of $<|B_L|>$ (the spatially averaged unsigned longitudinal flux in
the internetwork) were routinely being made and its variation with the spatial
resolution closely followed \citep{Almeida2003}. In a way, Hinode/SP results
have made this emphasis obsolete. We now know that these fields are largely
transverse and one should mainly care about $<|B_T|>$ or simply about the
spatially averaged $<|B|>$. We should caution here that these magnitudes are
obtained by observations of different Stokes parameters that have different
sensitivities to the real magnetic field components on the Sun and to the
fraction of the observed pixel that they occupy (the filling factor). Thus, the
steps to compute  $<|B|>$ from the observed $<|B_L|>$ and  $<|B_T|>$ are more
problematic than what one might anticipate. \cite{Lites2008} estimated that the
quiet sun map obtained by Hinode/SP had a $<|B_L^{app}|>$ of 11 G (or Mx
cm$^{-2}$) and a $<|B_T^{app}|>$ of 55 G. These estimates were based on using
integrals of the Stokes parameters that were calibrated against magnetic fluxes
but with no account for the fraction of the pixel occupied by the fields. This
is why they are named `apparent' fluxes. An analysis performed by
\cite{Orozco2007} of the same data, but this time using a Milne-Eddington (M-E)
inversion code including a filling factor as a free parameter, resulted also in
a predominantly transverse nature of the internetwork, albeit with a smaller
ratio of transverse to longitudinal apparent fluxes. 

It is no exaggeration to say that the ratio measured by
\cite{Lites2008} $<|B_T^{app}|>/<|B_L^{app}|>\sim 5$ came as a surprise
and was, thus, subjected to a deep scrutiny by the community. 

\begin{figure}
  \includegraphics[width=0.5\textwidth]{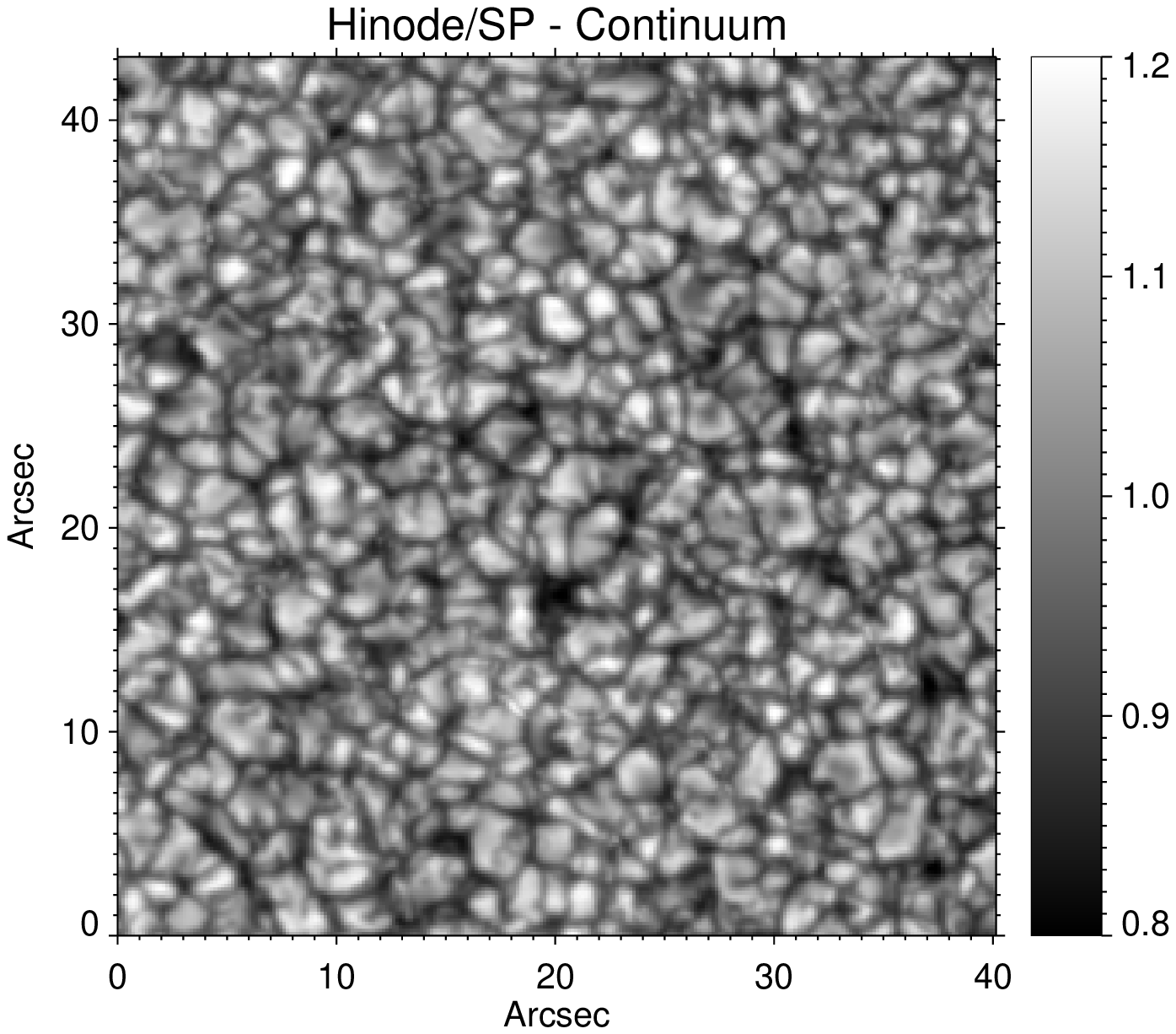}
  \includegraphics[width=0.5\textwidth]{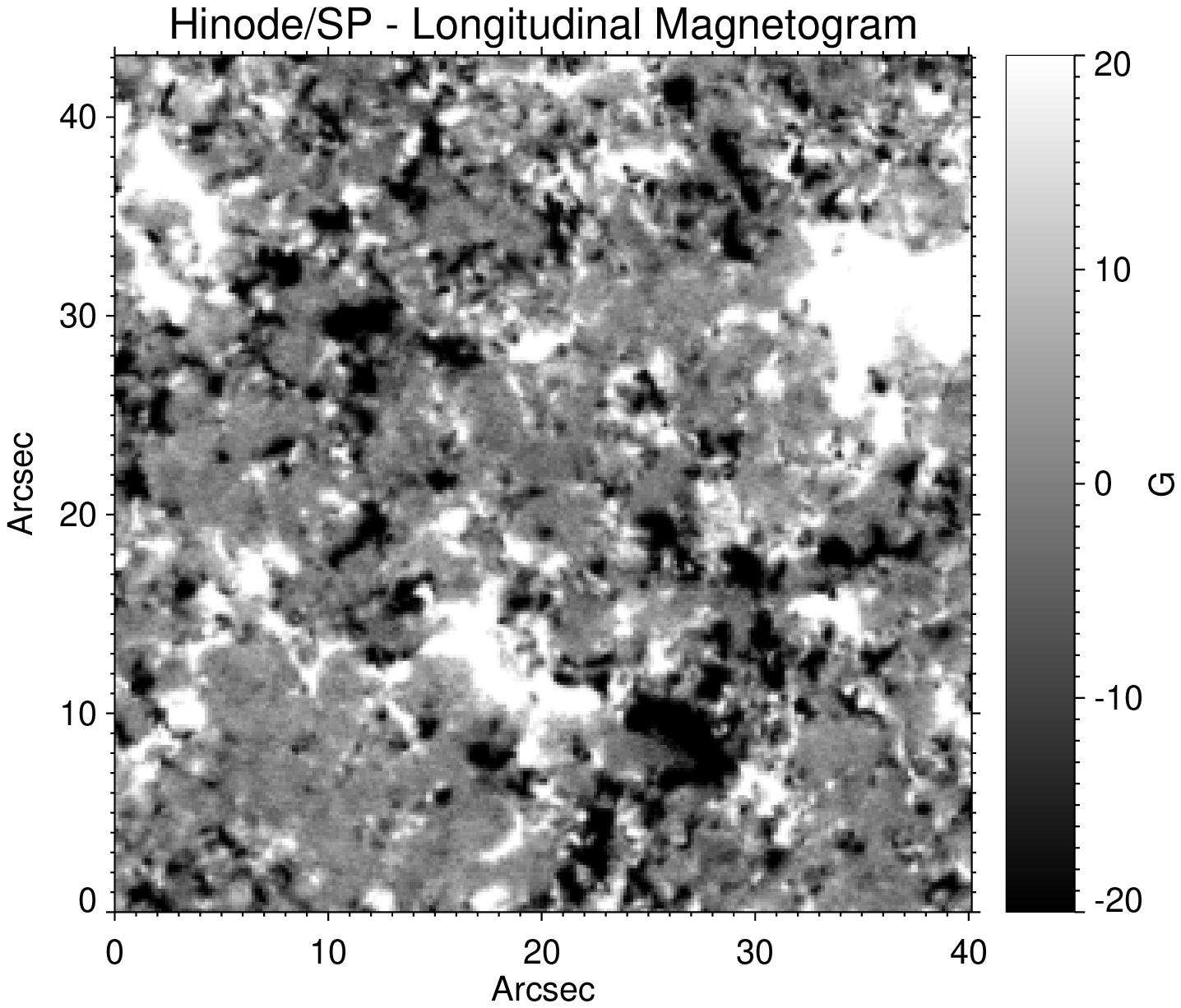}
  \includegraphics[width=0.5\textwidth]{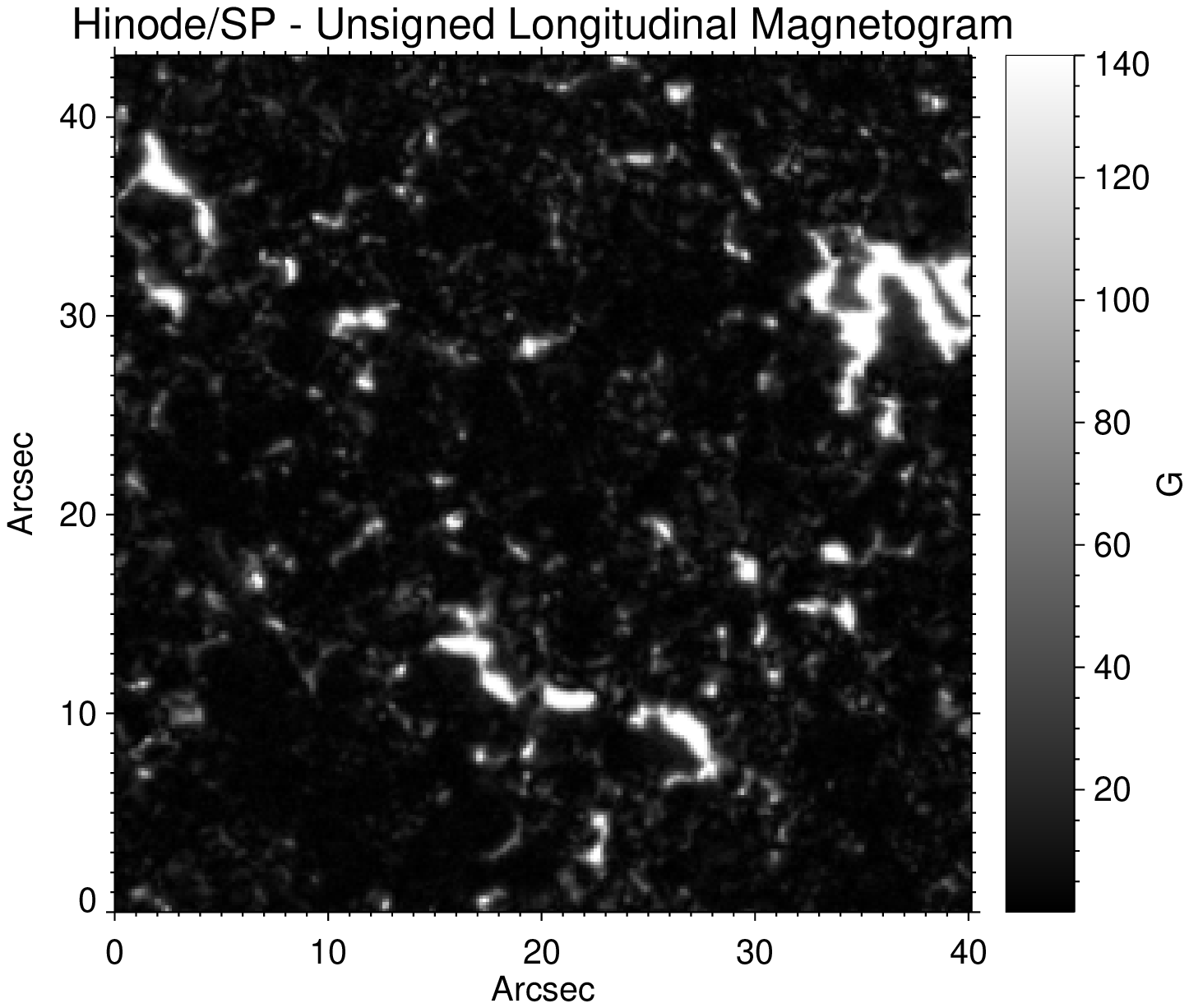}
  \includegraphics[width=0.5\textwidth]{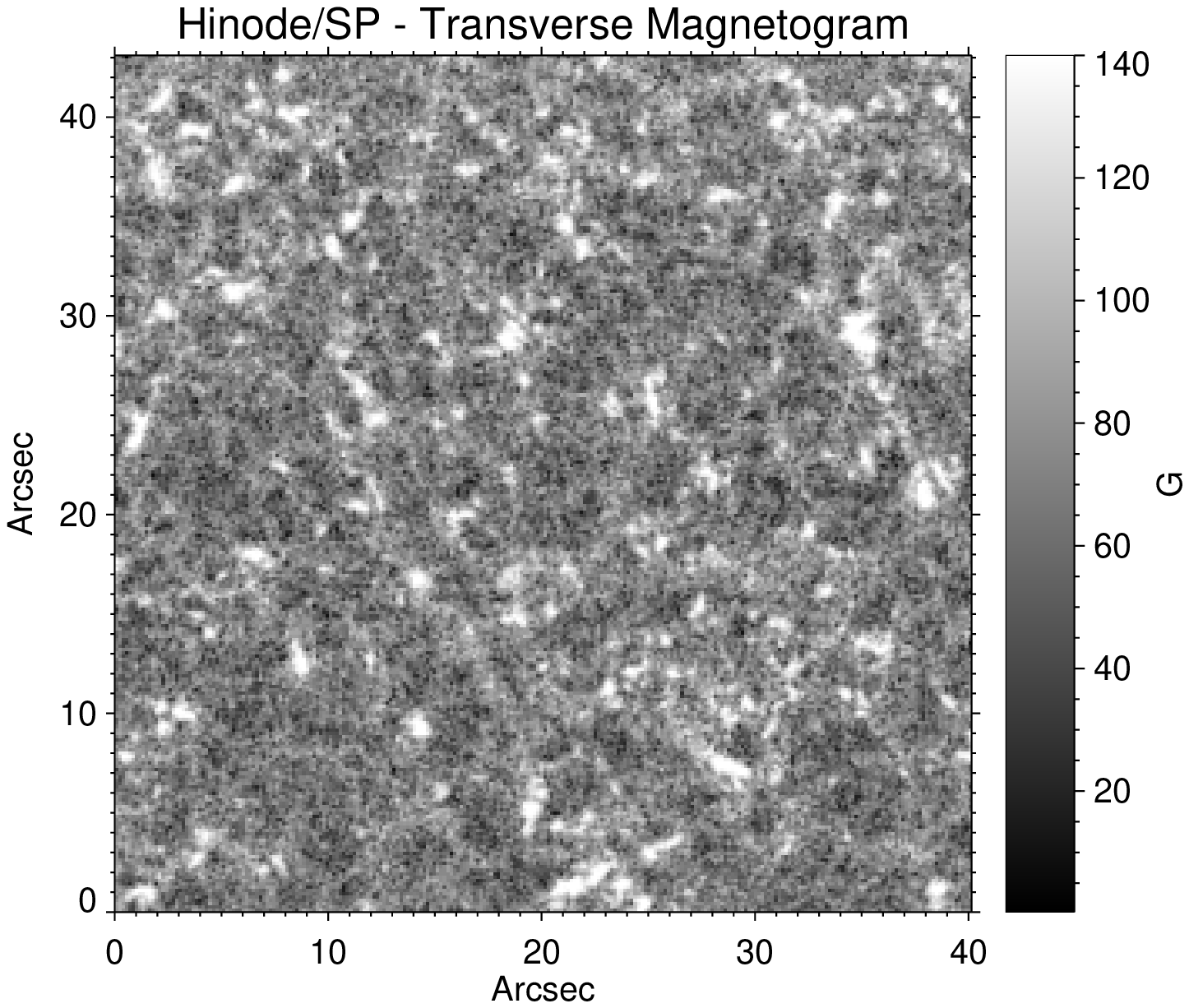}
\caption{Continuum intensity from the central portion of the \cite{Lites2008} Hinode/SP
map (top left), signed magnetogram from the same portion (top right), unsigned
magnetogram (bottom left) and transverse magnetogram (bottom right).
}
\label{fig:2}       
\end{figure}

Several factors can create a systematic bias in this ratio. Spatially averaging
noise affects a positively defined quantity such as $B_T$ in a way different
than what it does to a signed quantity ($B_L$). As the Zeeman effect has a
sensitivity different for each of these two components, the visibility of a
given field strength is different depending on whether it is a field aligned
with the LOS or perpendicular to it. In particular, fields close to the noise
levels translate into different visibility thresholds. Last but not least, there
is the already mentioned difference in how the filling factor couples to the
real field strengths and inclinations for each one of the two components and
depending on the specific method of analysis used. These effects and their
impact into the factor 5 obtained by the first Hinode/SP measurements have been
studied by various authors \citep[][see the latter for a review]{Asensio2009,
Borrero2011, Stenflo2011, Almeida2011, Steiner2012}.  It all translates into
understanding how exactly noise influences the final result given the method of
analysis one follows and the various thresholds for inclusion of a given pixel or
not. Depending on the specific case, different values for $<|B_T^{app}|>$,
$<|B_L^{app}|>$, or inverted parameters $B$, inclination, azimuth and filling
factor are obtained. Figure \ref{fig:2} shows the central portion of the same
magnetogram used in \cite{Lites2008} with the top panels displaying the
continuum intensity and $B_L^{app}$ scaled to $\pm 20$G.  Network bright points
are visible (e.g., top-right of the Figure) and the corresponding large Stokes
V signals evident in the magnetogram.  The bottom two images provided
$|B_L^{app}|$ (left) and $|B_T^{app}|$ (right) both scaled within {\em the same
range} [0,140]G. Almost all of the signals seen in the left image are
identified with network regions as can be recognized by inspection with the two
top panels. Thus, they will be excluded when computing $<|B_L^{app}|>$
for internetwork regions. In contrast, all of the signals seen in the right
panel correspond to internetwork and contribute to the $<|B_T^{app}|>$ average.
But note also in this last panel that regions with no apparent signals do not
show up as dark black as in the unsigned longitudinal image, but they show a
grey shade. This background is created by noise and it is larger than for the
longitudinal magnetogram simply because of the effects described above.  These
regions should neither be just included when computing $<|B_T^{app}|>$ nor
completely excluded (as some transverse fields might exist there). 
It is in all of these details that reside some of the
contradictory numbers that have been published. The most promising venue to
clarify the situation is that of reducing noise.  With the current
instrumentation this can only be done by using longer integrations. This 
strategy has been recently pursued very successfully by \cite{Orozco2012} and
\cite{Bellot2012}. They use slit integrations of 6.1 minutes that reach a
polarimetric sensitivity close to 10$^{-4}$, one order of magnitude better than
commonly achieved. For visible lines, this improvement lowers down the
detectability threshold for average transverse fields by a factor 2-3.  
\cite{Bellot2012} show that slit positions with almost
10 minute effective integration time that harbor linear polarization signals
basically everywhere (60\% above 4.5$\sigma$).  

The
increased exposure times impose a penalty in the sense that the spatial
resolution is decreased and evolutionary effects are intermingled in the final
results.

This should not give the
impression that these signals are always present on the Sun and detectable if
sufficient sensitivity is available as they might have
occurred in these pixels for only a fraction of the exposing time.
Nevertheless, the data obtained with these long exposures is perfectly suited
to reduced the noise induced bias present in the previous analysis of
Hinode/SP. The M-E inversions performed by \cite{Orozco2012} and
\cite{Bellot2012}, consistently show that the internetwork fields have
intrinsic field strength typically in the range of 100-200 G with basically no
kG present, a strongly peaked field inclination distribution near 90$^o$,
with most of the pixels displaying
inclinations in the range $[45^o,135^o]$, and an azimuth with no preferred
orientation. Filling factors move in the range of [0.2,0.4].  The authors  
inverted only those pixels that had a sufficiently large linear
polarization signal (in either Stokes Q and/or U) to ensure a reliable result
from the inversion. These results are largely free from most of the concerns
expressed so far on the nature of internetwork fields. One
criticism that remains to this analysis is the use of the M-E approximation and
inversions able to reproduce the asymmetries are desirable. But the main
conclusions from these recent analysis are likely to be confirmed by these more
complex inversions, as those based in the M-E approximation are known to
provide robust atmospheric means even in the presence of complex
stratifications \citep[see][]{Westendorp1998}. Thus internetwork fields do not
have an isotropic distribution of inclination as it has been argued in a number
of recent works \cite[][]{Asensio2009, Stenflo2011, Almeida2011} and their
field strengths are typically on the few hG range.

Inversion codes allow an inference of the filling factor as a separate free
parameter (admittedly, the most model dependent of all of them). Thus
\cite{Orozco2012} were able to give real mean (not apparent) fluxes.  For the
real fluxes, they obtained, $<|B_T|>\sim 198$G, $<|B_L|>\sim 64$G and
$<|B|>\sim 220$G (here the $<>$ average means those pixels with large enough
Stokes Q and U signals to allow a proper inversion, not the complete map).
Their ratio, now, is $<|B_T|>/<|B_L|>\sim 3.1$.  Note that because this ratio
has eliminated filling factor effects and the inclusion of only those points
that were inverted, it is not directly comparable to the number provided by
\cite{Lites2008}. However, it confirms the largely transverse nature 
of the internetwork fields as originally shown in that work. By eliminating
filling factor effects, this new ratio allows for a cleaner comparison with 
numerical simulations.

Interestingly, although not anticipated, numerical simulations of
magnetoconvection seem to have no problem in generating large amounts of
horizontal fields. Soon after the publication of the results from Hinode, the
SSDs simulations from \cite{Schussler2008} (MURaM code) and those from
\cite{Steiner2008} explained that these large amounts of transverse fields were
present in their simulation boxes at different heights. The simulations from
\cite{Steiner2008} were not dynamo simulations and used instead imposed fields
in both vertical and horizontal directions. These two initial
conditions generated a predominant horizontal field in the region where the FeI
line pair observed by Hinode forms.  Thus, while SSDs are capable of generating
a dominantly transverse field, it is not an exclusive property of them.  
In the work of \cite{Steiner2008}, it was
through the well known flux expulsion mechanism of vertical fields to the
intergranular lanes that horizontal field lines were expelled above the
granules in the overshooting region, generating  the predominant transverse fields. 
The quantitative comparison with the observed fields was
more complicated and both simulations fell short of the values observed.
\cite{Danilovic2010a} used the MURaM simulations and performed spectral
synthesis including instrument degradation and noise to compare the values
predicted from the simulations with those observed by \citep{Lites2008}. The
result was that while the factor five in the ratio of apparent mean fluxes came
naturally out of the SSD simulations, the absolute flux levels were close to
those observed only if the SSD fields were multiplied artificially be a factor
2-3. After this artificial increase, the average values in the SSD simulation
are closer to 100 G in the formation region of the FeI lines which is nicely
compatible with the peak field strength in \cite{Orozco2012}. Similarly, the
simulations presented by \cite{Steiner2008} resulted in average fields of the
order of 20 G for the transverse component and suffer from the same problem as
the SSD simulation. One could argue that the small ${\mathrm Re}_m$ numbers
achieved in the SSD simulations and the field strength introduced in the
simulations were too low and simply increasing them will explain the higher
fluxes  encountered by the observations. In any case, what was clearly established
from all these studies was the fact that the natural state of a magnetic field
component below equipartition strengths and closely coupled with the solar
granulation is that of a predominant transverse field component as found by
Hinode/SP.  Another conclusion is that while SSDs are compatible with this
result, the latter cannot be offered as a demonstration of their existence at
the solar surface as a non-dynamo magnetoconvective simulation found the same
results.

Another attempt to investigate whether the internetwork fields are generated by
dynamo action has been presented recently by \cite{Lites2011}. In this work,
the polarity imbalance of the internetwork field regions in 45 Hinode/SP maps
is studied. Arguably, polarity balance is considered a necessary outcome of an
SSD. However, and as stated by \cite{Lites2011}, this is not a sufficient
condition as after a sufficiently large number of turnover times, the same
polarity balance is to be expected in non-dynamo simulations. The difficulty
here stems from the fact that to measure the internetwork polarity imbalance one
needs to carefully isolate these fields from network ones. If the internetwork
fields originate somehow by the shredding of nearby network fields, one expects
the internetwork to have the same sign of the imbalance and a proportionality
between the two. Interestingly enough, while no scaling with the unsigned flux
was found, a suggestive correlation between the signed flux imbalance in the
internetwork and that in the nearby network was measured (see Figure 3 of the
paper).  However, being conclusive with this result is difficult as the
isolation between internetwork and network fields is always problematic. This
approach deserves further study, probably including some stray-light correction
that deconvolves the wings of the spread function. This would allow to
decontaminate the internetwork fields from the surrounding network contribution
and allow a more reliable study of the resulting polarity imbalances.  

It is clear that it will be very difficult to conclusively demonstrate that the
hG, predominantly transverse internetwork fields originate from a granulation
driven SSD. As mentioned in the above paragraph, we can only aim at disproving
the SSD hypothesis rather than expect a firm confirmation of its presence. In
Section 4, we present a recent result that, if consolidated, could be considered
as one such refutation.

\subsection{Hanle signals}

Scattering polarization in spectral lines and its modification via the Hanle
effect allows to study a completely different parameter space of solar
magnetism not accessible with the Zeeman effect \cite[see the reviews
in][]{Trujillo2006, dewijn2009, Stenflo2011}.  A tangled field distribution
within the resolution element is invisible through the Zeeman effect but can
leave a clear imprint in the linear polarization profiles as long as the field
strengths are below the Hanle saturation values (typically, a few hundred Gauss). 
While strong homogeneous vertical fields are hardly hidden in Stokes V through the
Zeeman effect, weak disorganized transverse fields show up easily in Stokes Q
and U of selected spectral lines thanks to the Hanle effect.  This different
sensitivity of Hanle effect has provided recently \citep{Trujillo2011} a
seemingly alternative description of the Quiet Sun fields to that discussed
above (and inferred from the Zeeman effect).  Ever since the early studies
\citep[]{Stenflo1982},  the Hanle signals have been interpreted as being due to a
tangled mixed polarity field. For this reason, it has always been natural to
associate these fields with the outcome of a turbulent dynamo \citep[see,
e.g.][]{Vogler2007, Pieta2010} This alternative description can be summarized
as follows.  The center-to-limb variation of the depolarization signals
observed in the optically thick SrI 4607 \AA~line suggests the presence of a
tangled field with characteristics strengths of $B\sim<130>$ G (as mentioned in
the Introduction). This value is computed using realistic atmospheric models of
the solar surface and complex 3D radiative transfer
calculations\citep{Trujillo2004, Shchu2011}.  In contrast, the analysis of the
depolarization signals of a set of optically thin molecular lines suggests a
mean field of the order of only $<10>$ G \citep{Trujillo2004, Kleint2011}.
These two distinct fields can be made compatible if one realizes that the
molecular lines are entirely formed in the hot smooth upflowing granules and
that these lines are blind to the fields present in the turbulent intergranular
lanes \citep{Trujillo2004}. The SrI line however sees the fields in both regions
and, in order to give rise to average values of 130 G over the whole solar
surface, \cite{Trujillo2004} concluded that one must have $<B> \apgt 200$ G
(very close to the Hanle saturation regime for this line) inside the lanes.
This intergranular fields would generate most of the SrI depolarization while
having no effect in the linear polarization signals from the molecular lines.
It is interesting to point out that when the SrI depolarization levels are
computed with the MURaM SSD simulations, \cite{Shchu2011} find that the
depolarization levels are far too low. This
is not surprising as these SSD simulations reach fields of the order of 20-30 G,
very far from the typical 130 G that is needed. The only way in which an agreement
with the observed depolarization values could be achieved was by multiplying
everywhere the field strengths in the simulations by a factor 12. This factor
is four times larger than that needed by \cite{Danilovic2010a} to match the
observed apparent mean Zeeman fluxes of Hinode/SP with those in the MURaM
simulations.

This description of the quiet sun fields as inferred from the Hanle signals is
generally accepted and no clear alternative exist. However, it is clear that
one would like to see it confirmed by an analysis that is less model dependent
\citep[see][who point out that the analysis made using molecular lines is
differential in contrast to that made with the atomic lines]{Kleint2011,
Stenflo2012}. The fact that the SrI line is so close to the saturation regime has
also brought some criticism of the actual interpretation of the observed
depolarization levels \citep{Almeida2005}. Indeed, very little variation of the
SrI polarization signals has been detected and this can be used to argue
that the tangled hidden field is independent of the solar cycle, favoring an
origin related to SSD action \citep{Trujillo2004, Vogler2007, Pieta2010}.  One
other explanation, of course, could be that the signals in the intergranular
lanes change with the activity cycle but we do not see the effect because they
are always in the Hanle saturation regime.

In any case, the existence of this turbulent unresolved field is well
established and one would like to understand if it bears any relation with the
internetwork fields observed by Hinode/SP and described before. Note that when
we say unresolved here, we refer to the Hanle observations used in the analysis
that were obtained over large spatial scales (several arcseconds) and with
exposure times of the order of one minute or so. Taken at face value, what the
Hanle measurements need are small scale (arcsecond scale or below) non-vertical
field patches (the Hanle effect is insensitive to vertical fields),
with field strengths below equipartition with granulation ($\sim 400$ G).  It
is evident that these are the properties of the internetwork fields described
by \cite{Bellot2012} and \cite{Orozco2012} and one is tempted to conclude that
the Hinode/SP fields are also the fields corresponding to the Hanle signals.
\cite{Bellot2012} specifically mentions this possibility.  One would be tempted
to go further and state that it would be rather strange to have the Sun harboring
two families of fields with so many things in common but that
are totally unrelated.  Thus, we also favor here the identification of the
Hinode/SP internetwork fields with those that produce the Hanle depolarization
signals \cite[see also][]{Lites2009}.  Perhaps, the only difficulty we
encounter with this identification is the well known fact that the HIFs (the
internetwork transverse fields) have a clear preference to be located at the
borders of granules \citep{Lites2008, Danilovic2010b} not in the
intergranular lanes as required by the SrI depolarization measurements.
However, this could be a minor problem as a granular border might be
sufficiently close to what is needed. An evident way to test this
identification would be to invert a volume of the Hinode/SP internetwork
observations with an inversion code that provides the complete atmospheric
stratification such as the SIR code \citep{SIR1992}. Then, perform the Hanle
synthesis as in \cite{Trujillo2004} and \cite{Kleint2011} for the atomic
and molecular lines and compare the result with the observations. In this
exercise some assumptions about the upper layers might be needed to extend the
retrieved atmospheres over the range of formation of SrI line, but such an
extension can be reasonably done. 

In any event, it is clear that spatially resolved Hanle depolarization
measurements \citep{Stenflo2012} that tell us where exactly the SrI
depolarization signals occur in the Sun are urgent. 

\section{Deep magnetograms and `dead' calm areas: implications}
\label{sec:4}

\begin{figure}
  \includegraphics[width=1.\textwidth]{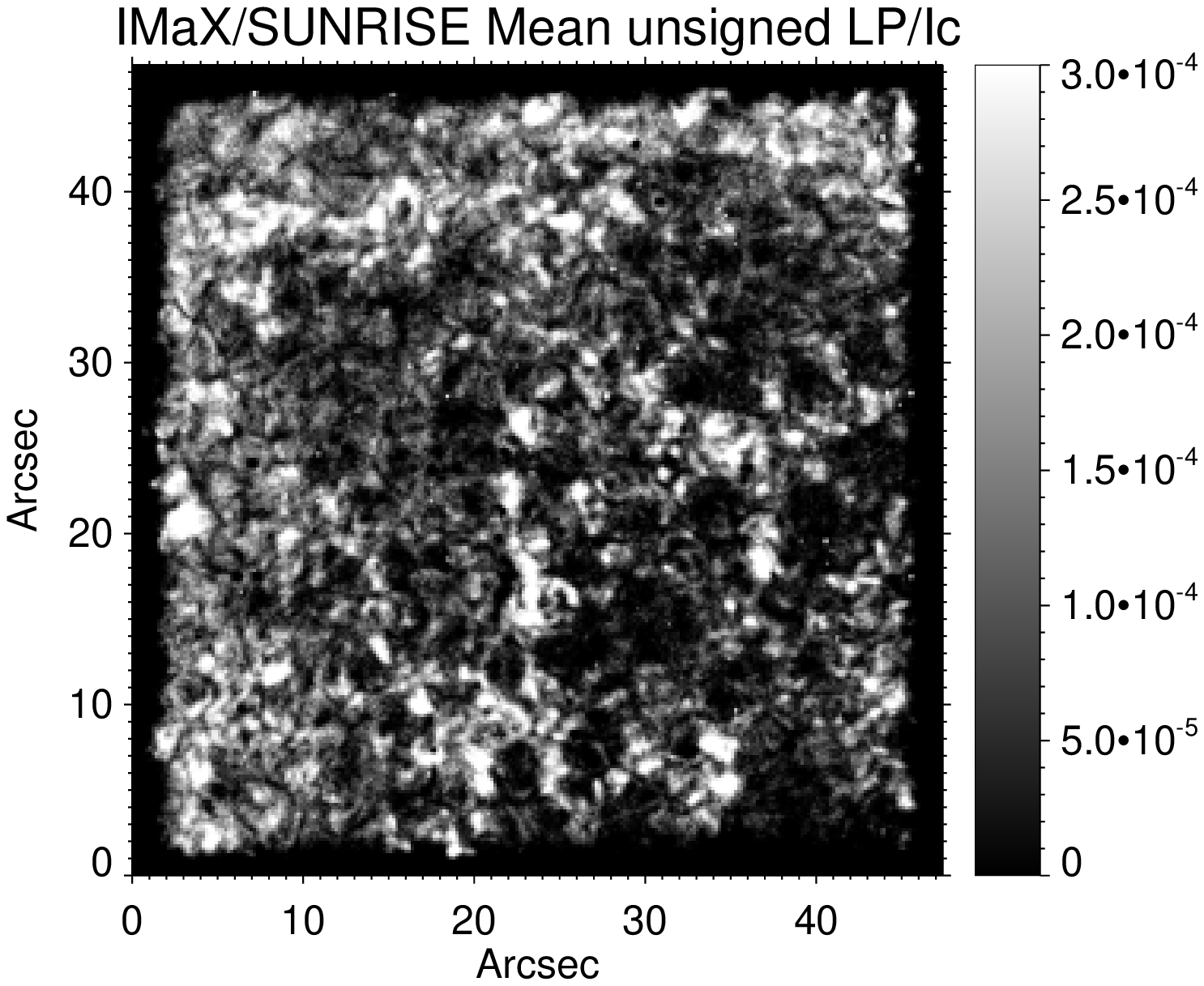}
  \includegraphics[width=1.\textwidth]{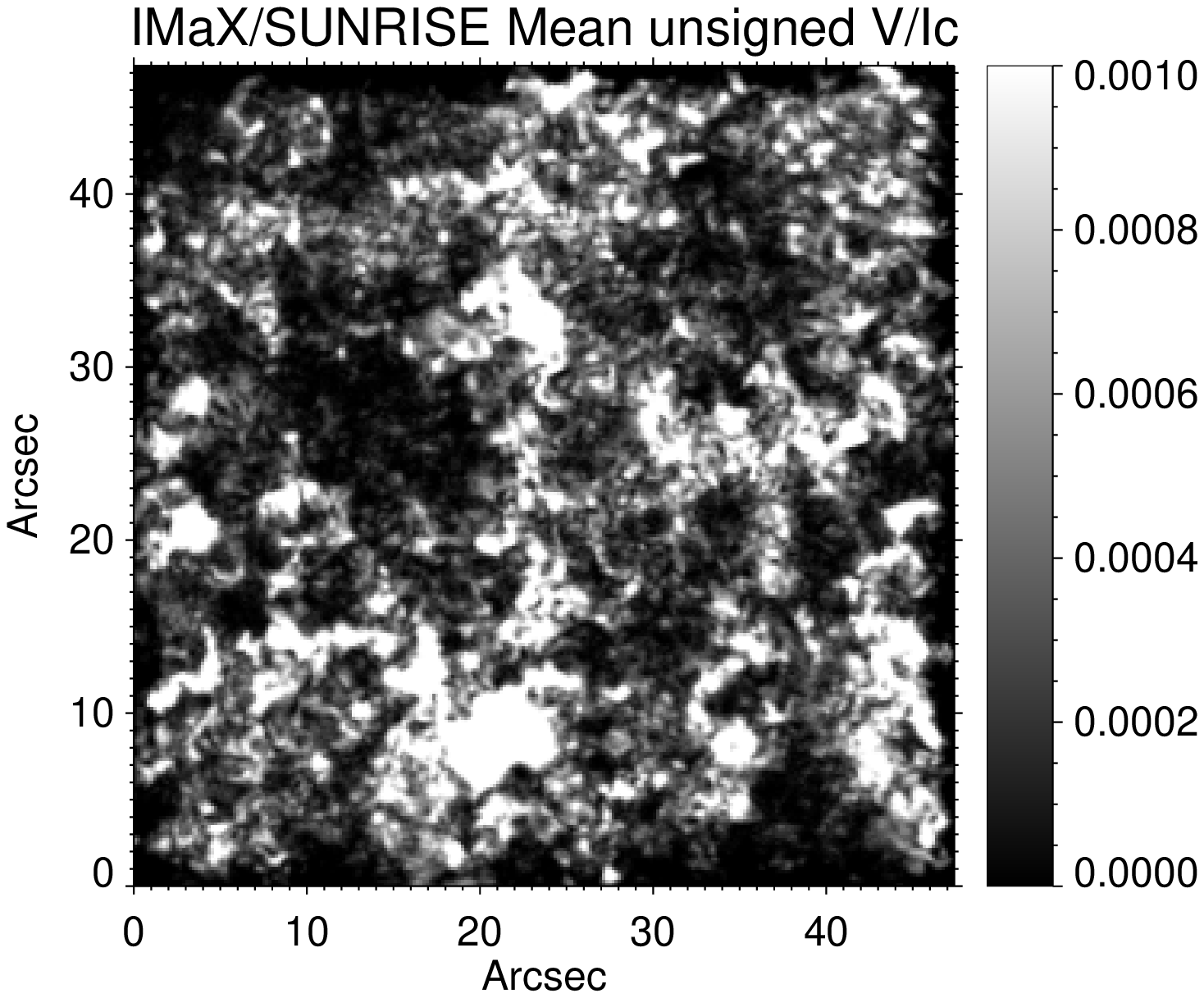}
\caption{
SUNRISE/IMaX deep linear polarization magnetogram integrated
over a 30 minute period (top). 
SUNRISE/IMaX deep circular  polarization magnetogram integrated
over the same period (bottom). Both are computed according
to the definitions in Eq. 1.
}
\label{fig:3}       
\end{figure}

As commented in Section \ref{sec:1}, dynamo action concentrates in the
turbulent downflows as field line stretching and amplification is more
efficient there.  A dramatic visualization of this can be seen in Figure 1 of
\cite{Schussler2008}. Basically all of the intergranular lanes are seen to
participate in this dynamo swing.  One can then expect that the observable
effects of an SSD driven by the granulation would be distributed homogeneously
over spatial scales similar to that of the granulation after a period of
time long compared with the lifetime of an intergranular lane \cite[which is
similar to that of the granules themselves, 10 minutes, cf.][]{Title1989}.  In
particular, the internetwork fields seen by Hinode/SP must be uniformly
distributed over granulation scales.  Thus, if we are able to
somehow make a statistics of the location of these fields, the inferred spatial
distribution must reflect the spatial scale of the granulation and have no
voids larger than the typical size of one granule (or at least, the probability
of occurrence of such voids will be small, see below). On the contrary, if one
finds areas on the quiet sun where many intergranular lanes have existed but
none of the expected effects of a presumed granular SSD are seen, one can
conclude that no such SSD has been detected.

While snapshots from SSDs have been used to make spectral synthesis including
the degradation effects of telescopes and detectors \citep{Danilovic2010a,
Shchu2011}, what one needs is a complete time series that allows the study of
the generation and disappearance of the dynamo fields at current spatial
resolutions and polarimetric sensitivities. This study is unfortunately not
available yet. From an observational perspective, we have seen how internetwork
fields evolve at the solar surface with unprecedented detail.  Both Hinode/SP
and SUNRISE/IMaX have convincingly shown that internetwork fields are fed into
the surface in the form of emerging small scale loops (rather than in the form
of, for example, spontaneous appearances of newly created flux patches in the
intergranular lanes). Hinode/SP with its superior spectral resolution and
coverage has allowed to study a large number of such events \cite[see][the
latter for ground observations in the infrared]{Rebeca2007, Marian2009,
Ishikawa2010, Viticchie2012, Gomory2010} with great detail. Basically a
horizontal patch is first detected in linear polarization that later displays
two opposite polarity footpoints that move apart. The distance between the
footpoints of these small scale loops
is typically 1 Mm, the lifetime 10 minutes and the magnetic flux of
around $10^{17}$ Mx \cite[see the statistic in][]{Marian2009}.
\cite{Danilovic2010b}, using data from SUNRISE/IMaX, identified a large number
(thousands) of HIFs that they associate with flux emergence in the form of
loops. Such a large amount of occurrences emphasize this process as the main
source for internetwork flux at present resolutions and sensitivities. A case
in point is that described by \cite{Salvo2012} who analyzes in detail what is
probably the largest quiet sun loop ever observed. In this case, a maximum
footpoint separation of 4.5 Mm is achieved with a magnetic flux content of 6
$10^{17}$ Mx and a duration of 25 minutes. This quiet sun bipole, that has one
order of magnitude less flux than the smallest ephemeral region studied by
\cite{Hagenaar2003}, is arguably not generated by any process that occurs at
granular scales.  It is known that dynamo simulations generate similar
loop-like structures (as the horizontal fields described in Section \ref{sec:2}
are part of them). But a comparative study of the maximum footpoint separation,
flux content, etc., is missing. 

A recent study about loop emergence in the quiet sun has recently been
published that is relevant for our discussion.  Using the two time series of
the first flight day obtained with the IMaX instrument, \cite{Marian2012}
studied the emergence of 497 magnetic loops identified in them. They estimate
an event rate of 0.25 loop h$^{-1}$ arcsec$^{-2}$.  If we associate a typical
linear size of 2 arcsec for a granule and a lifetime of 10 minutes, this
rate can be translated into 0.17 loops per granule (it takes 6 granules to get
one quiet sun loop). Each one of these time series lasted for about 30 minutes,
so granulation was efficiently created and destroyed over their timespan. Such
a large number of detected loops allowed them to study the spatial distribution
of these magnetic flux emergence processes. The result they obtained was that
the spatial distribution of loop events was far from homogeneous at granular
scales.  They found what they termed "dead calm" areas where simply no loop was
seen to emerge during the time series.  One could argue that these calm areas
can be created by chance and that their existence is simply a mere coincidence.
However, the authors perform a statistical study of the likelihood of such
voids given their size (70-100 arcsec$^2$) under the assumption of a spatially
uniform loop emergence probability. They modeled the probability of finding
one such large circular void with a resulting estimate of 3$\times 10^{-4}$.
Two such dead calm areas were cleanly identified. 

Let us show from another perspective how unlikely this result is.  In voids of
this size, one can fit around 20 typical granules at any given time. As the
time series covered 3 granular lifetimes, 60 granules existed inside them which
would have given rise to at least 10 loops at the above rate of creation, but
none was found. And it occurred in two unrelated regions. How is it possible
that if magnetic loop emergence is the observable outcome of SSD action, one
encounters regions where this is not activated? While we find this result
highly incompatible with the existence of a granularly driven SSD, we need a
solid comparison with simulations including all of the possible
observational biases to provide a firm answer.

We must note that a non-uniform distribution of quiet sun fields has been known
for some time. They were identified by \cite[][and references
therein]{Lites2008}, who pointed out the existence of mesogranular scale voids
in the Hinode/SP map. The preference of internetwork fields to be located at
mesogranular scales has been clearly demonstrated using the same SUNRISE/IMaX
data as that used to identify the voids \cite[see][]{Lotfi2011}, but note that
these voids were much larger than a mesogranule and have a scale closer to that
of a small supergranule \cite[see, e.g.][who give radius in the range of 8 to
30 arcsec]{Meunier2007}.

One could argue that whereas no loop emerged in these dead calm areas, they
were probably not devoid of some subtle form of internetwork field presence.
However, inspection of the IMaX data with a scaling close to the noise levels
readily shows that the locations of these voids clearly harbored less activity
than the rest of the observed area. To prove this point more clearly, we produced
two time averages of the IMaX time series that are shown in Fig. \ref{fig:3}.
The two quantities that are displayed correspond to deep magnetograms computed
as: 
\begin{equation} \overline{LP} ={{1}\over{N}}\sum_{i=1}^N \Bigl
({{\sqrt{Q_i^2+U_i^2}}\over{I_c}}-\epsilon_{LP}\Bigr )~~~~~~ \overline V
={{1}\over{N}}\sum_{i=1}^N \Bigl ({{|V_i|}\over{I_c}}-\epsilon_V\Bigr )
\end{equation} 
where $\epsilon_V$ and $\epsilon_{LP}$ are quantities inferred
from the data that allow convenient reduction of the noise when doing the time
averages of the otherwise positively defined quantities.  The deep magnetograms
in Fig. \ref{fig:3} evidence the same two voids as detected by
\cite{Marian2012}.  They are centred at coordinates [15,25] and [35,37]. The
voids are visible in both the linear and the circular polarization deep
magnetograms. Note that the scaling of both magnetograms saturates at $3\times
10^{-4}$ for $\overline{LP}$ and at $10^{-3}$ for $\overline{V}$. The
calibration constants published by \cite{Pillet2011} would have translated
these values into 45 and 5 G respectively. Measurements below these fluxes are
at the limit of state of the art imaging magnetographs. These deep magnetograms
show that they had reduced levels of magnetic activity and it is correct to
refer to them as magnetically calm (maybe, not dead).  The question, of course, is
what was special about the many intergranular lanes that populated these
regions that prevented them from displaying the magnetic activity levels seen
elsewhere.

\section{Where do we go from here?: higher sensitivities and higher latitudes}
\label{sec:5}

The simulations described in Section \ref{sec:1} have reached ${\mathrm
Re}_m\in [5000,8000]$ \citep{Pieta2009a}. However, they fail to provide the
magnetization levels needed to explain, both, the Hanle depolarization signals
of the SrI line and the Hinode/SP fluxes \citep[see][]{Danilovic2010a,
Shchu2011, Orozco2012}.  This is often explained arguing that the simulations
are still far from the magnetic Reynolds numbers of the Sun. But the values
that are achieved by them are {\em only} about one order of
magnitude below the expected values at the solar surface. In contrast, they
reach, at best, similar ${\mathrm Re}$ values, which are $10^7$ times smaller
than what we encounter on the Sun (see Fig. \ref{fig:1}, where the values
achieved by the simulations are marked by an X near the surface).  There are
about 5 orders of magnitudes in the inertial range below the resistive scale,
$l_\eta$, populated with cells that generate an enhanced turbulent diffusion
that are not present in the simulations. As mentioned
in Section \ref{sec:1}, recent numerical simulations show that this might not be
a problem to obtain dynamo action in the low ${\mathrm P}_m$ regime.
But they also tell us that the value of the critical magnetic Reynolds number
needed ${\mathrm Re}_m^C$ to sustain dynamo action increases sharply when
${\mathrm P}_m <1$.  A factor 3-7 increase in the value of ${\mathrm Re}_m^C$ is
expected.  As in the MURaM simulations this number is $\sim 2000$, we expect
this code to display a dynamo only when ${\mathrm Re}_m >6000$ or larger as
soon as they use Prandtl numbers in the right ballpark of the problem.  Let us
see the implications for some of the inferences that are made using the
available numerical simulations.  In the case of the MURaM runs,
\cite{Danilovic2010a} found a plausible scaling of the saturated field
strengths with $\sim {\mathrm Re}_m^{1/2}$. Their run G with ${\mathrm
Re}_m\sim 5200$ has a mean field of 30 G (at $\log\tau \approx -1$). If we
scale it to a solar value of ${\mathrm Re}_m\sim 10^5$ following this square
root scaling, we obtain 130 G. This is the value needed to explain the Hanle
depolarization measurements of the SrI line \cite[and the most probable field
strength found in the internetwork by][]{Orozco2012}. This nice agreement was
already pointed out by \cite{Shchu2011}. However, for the reasons explained
above, this is probably a mere coincidence. If this simulation would have been
done with the same ${\mathrm Re}_m$ but with ${\mathrm P}_m$ of, say 0.1 (as in
the simulations proving dynamo action at low Prandtl numbers),  no dynamo
action would have been found.  The field strength to introduce in the above
scaling would have been 0 G instead of 30 G. 

In discussing the accepted view about the existence of dynamo action at the
solar surface, we have gone a step further and suggested that the observed
spatial distribution of quiet sun fields seems to be at odds with a granularly
driven SSD. The argument used to make this claim was that in such a dynamo, all
the magnetic byproducts must necessarily have a uniform spatial distribution at
scales above that of a granule. We have found, however, that there are voids of
magnetic activity (dead calm areas) in, both, the average apparent longitudinal
and transverse fluxes of deep magnetograms and in the distribution of emerging
loops.  In fact, we have translated the loop emergence rate found by
\cite{Marian2012} into a rate of 1 loop per 6 granules which can also be
thought of as 1 emerging loop per mesogranule. It is clear that mesogranulation
scales are very relevant for the quiet sun fields. SSD simulations including
them are needed to see if they explain the presence of these voids in magnetic
activity and emerging loop frequency. Given the fact that as we go deeper into
the Sun, one reaches higher ${\mathrm Re}_m$ and ${\mathrm P}_m$ values, it is
very tempting to suggest that SSDs acting in a range of convective scales
larger than the granular ones are those that give rise to the presently observed
internetwork fields. Note that in his work, \cite{Cattaneo1999} already
mentioned larger scales, such as those of the supergranulation, as possible
places where to host dynamo activity.  This might still be compatible with an
SSD at granular scales that generates fields much weaker and that have not yet
been observed by any of our currently available diagnostic techniques. 

We have also suggested that the internetwork fields revealed by Hinode/SP and
those that generate the Hanle depolarization of some atomic lines might have a
lot more in common than previously thought \citep[see however][]{Lites2009,
Bellot2012}. The reason for this identification is based on the results from
\cite{Orozco2012} and \cite{Bellot2012} who have shown that the quiet sun fields
share the same field strengths, inclinations and azimuths than those needed by
the fields detected with the Hanle effect.  This is, of course, compatible with
a continuous spectrum of fields in which the ranges corresponding to the two types
of fields (the internetwork and the `hidden' Hanle fields) simply overlap over
a much larger fraction than thought so far. Of course, the internetwork fields
will also include a fraction of vertical kG fields that contributes nothing to
the Hanle depolarization and the weak granular fields that depolarize the light
in molecular lines never make an imprint in the FeI Zeeman lines observed
by Hinode/SP at current sensitivities.  But, of this continuous distribution of
fields, the range of hG strengths with largely transverse orientations
and spatially organized at granular scales, contributes simultaneously to, both,
the Zemman and the Hanle observations. If this result is confirmed (as, e.g., with the SIR
inversions and the Hanle computations mentioned in Section \ref{sec:3}) an
important step to clarify the currently complex debate of the nature of the
quiet sun magnetism would be achieved.  We also want to stress that 
the evidence that
the internetwork field component is composed of largely transverse fields
renders the debate about the mean value of $<|B^{app}_L|>$ (or of $<|B_L|>$ for
that matter) obsolete. Stokes V (longitudinal) magnetograms of the internetwork 
simply show a rather incomplete picture of these fields.

Note that many of the results commented above have benefited from high
polarimetric sensitivities. Hanle measurements have always been very demanding
in polarimetric accuracy. The deep magnetograms of SUNRISE/IMaX and the long
integrations with Hinode/SP used by \cite{Orozco2012} both were at the
$10^{-4}$ polarimetric sensitivity. This is not a coincidence. Much of the
future progress will be achieved with sensitivities in this range. Those more
regularly reached in present day observations, $10^{-3}$, are due to
instrumental limitations that have nothing to do with physical processes in the
Sun. Polarimeters observing the solar photosphere with sensitivities of
$10^{-4}$ and sub-arcsecond resolutions using the Zeeman and the Hanle effects
will consolidate (or refute) many of the aspects commented here.  The need for
high spatial resolution observations of the SrI depolarization cannot be emphasized
enough \citep{Stenflo2012}. These targets demand large apertures similar
to those planned for future facilities such as the 4m class ground-based
telescopes \cite[ATST, EST][]{Keil2011, Collados2010} and the Japanese led
Solar-C mission (1.5m aperture). All these facilities will likely have to be
used outside of the diffraction limit to pursue high sensitivity spectropolarimetry
thanks to their large collecting areas \cite{Keller1999}.

Finally, observing regions of the Sun hardly reachable from the ecliptic will
also help to clarify the nature of the quiet sun magnetism. The ESA-led Solar
Orbiter mission \citep{Muller2012} will carry on-board a magnetograph
\citep[the Polarimetric and Helioseismic Imager, PHI;][]{Gandorfer2011} similar
to that of SDO/HMI \citep[][]{Scherrer2012} and will observe the dynamics of
the solar poles from an inclination of 35$^{o}$ with respect to the solar
equator. Observing the poles is crucial in this discussion because they
represent the regions at farther distances from the activity belts in the Sun.
In the absence of an SSD mechanism working at the solar surface, the origin of
the internetwork fields can only be explained as a result of the cascading down
towards the smallest scales \citep[see the discussion in][]{Schussler2008} of
the global dynamo fields. This effect is inevitably present on the Sun, but
whether it affects only to network fields or to internetwork ones can best be discerned
by observing their latitudinal properties with good spatial resolution
and sensitivity.

\begin{acknowledgements}
This work has been partially funded by the Spanish
MINECO through Project No. AYA200AYA2011-29833-C06. 
Comments on an original version of the manuscript by
D. Orozco, K. Petrovay and an unknown referee are gratefully acknowledged.
ISSI support to attend the meeting is also acknowledged. 
\end{acknowledgements}

\bibliographystyle{aps-nameyear}      
\bibliography{mybib_2}   

\begin{thebibliography}{75}
\ifx \bisbn   \undefined \def \bisbn  #1{ISBN #1}\fi
\ifx \binits  \undefined \def \binits#1{#1} \fi
\ifx \bauthor  \undefined \def \bauthor#1{#1} \fi
\ifx \bjtitle  \undefined \def \bjtitle#1{\textrm{#1}}\fi
\ifx \batitle  \undefined \def \batitle#1{#1} \fi
\ifx \bctitle  \undefined \def \bctitle#1{#1} \fi
\ifx \bvolume  \undefined \def \bvolume#1{\textbf{#1}}\fi
\ifx \byear  \undefined \def \byear#1{#1} \fi
\ifx \bissue  \undefined \def \bissue#1{#1} \fi
\ifx \bfpage  \undefined \def \bfpage#1{#1} \fi
\ifx \blpage  \undefined \def \blpage #1{#1} \fi
\ifx \burl  \undefined \def \burl#1{#1} \fi
\ifx \doiurl  \undefined \def \doiurl#1{#1} \fi
\ifx \betal  \undefined \def \betal{et al.} \fi
\ifx \binstitute  \undefined \def \binstitute#1{#1} \fi
\ifx \beditor  \undefined \def \beditor#1{#1} \fi
\ifx \bpublisher  \undefined \def \bpublisher#1{#1} \fi
\ifx \bbtitle  \undefined \def \bbtitle#1{\textit{#1}} \fi
\ifx \bedition  \undefined \def \bedition#1{#1} \fi
\ifx \bseriesno  \undefined \def \bseriesno#1{#1} \fi
\ifx \blocation  \undefined \def \blocation#1{#1} \fi
\ifx \bsertitle  \undefined \def \bsertitle#1{#1} \fi
\ifx \bsnm \undefined \def \bsnm#1{#1} \fi
\ifx \bsuffix \undefined \def \bsuffix#1{#1} \fi
\ifx \bparticle \undefined \def \bparticle#1{#1} \fi
\ifx \barticle \undefined \def \barticle#1{#1} \fi
\ifx \botherref \undefined \def \botherref #1{#1} \fi
\ifx \url \undefined \def \url#1{#1} \fi
\ifx \bchapter \undefined \def \bchapter#1{#1} \fi
\ifx \bbook \undefined \def \bbook#1{#1} \fi
\ifx \bcomment \undefined \def \bcomment#1{#1} \fi
\ifx \oauthor \undefined \def \oauthor#1{#1} \fi
\ifx \citeauthoryear \undefined \def \citeauthoryear#1{#1} \fi
\ifx \texttildelow  \undefined \def \texttildelow{\symbol{126}} \fi
\def \endbibitem {}
\ifx \bconflocation  \undefined \def \bconflocation#1{#1} \fi

\bibitem[\protect\citeauthoryear{{Abbett}}{2007}]{Abbett2007}
\begin{barticle}
\bauthor{\binits{W.P.} \bsnm{{Abbett}}},
\batitle{{The Magnetic Connection between the Convection Zone and Corona in the
  Quiet Sun}}.
\bjtitle{\apj}
\bvolume{665},
\bfpage{1469}--\blpage{1488}
(\byear{2007})
\end{barticle}
\endbibitem

\bibitem[\protect\citeauthoryear{{Asensio Ramos}}{2009}]{Asensio2009}
\begin{barticle}
\bauthor{\binits{A.} \bsnm{{Asensio Ramos}}},
\batitle{{Evidence for Quasi-Isotropic Magnetic Fields from Hinode Quiet-Sun
  Observations}}.
\bjtitle{\apj}
\bvolume{701},
\bfpage{1032}--\blpage{1043}
(\byear{2009})
\end{barticle}
\endbibitem

\bibitem[\protect\citeauthoryear{{Batchelor}}{1950}]{Batchelor1950}
\begin{barticle}
\bauthor{\binits{G.K.} \bsnm{{Batchelor}}},
\batitle{{On the Spontaneous Magnetic Field in a Conducting Liquid in Turbulent
  Motion}}.
\bjtitle{Royal Society of London Proceedings Series A}
\bvolume{201},
\bfpage{405}--\blpage{416}
(\byear{1950})
\end{barticle}
\endbibitem

\bibitem[\protect\citeauthoryear{{Bellot Rubio} and {Orozco
  Su{\'a}rez}}{2012}]{Bellot2012}
\begin{barticle}
\bauthor{\binits{L.R.} \bsnm{{Bellot Rubio}}},
\bauthor{\binits{D.} \bsnm{{Orozco Su{\'a}rez}}},
\batitle{{Pervasive Linear Polarization Signals in the Quiet Sun}}.
\bjtitle{\apj}
\bvolume{757},
\bfpage{19}
(\byear{2012})
\end{barticle}
\endbibitem

\bibitem[\protect\citeauthoryear{{Boldyrev} and
  {Cattaneo}}{2004}]{Boldryrev2004}
\begin{barticle}
\bauthor{\binits{S.} \bsnm{{Boldyrev}}},
\bauthor{\binits{F.} \bsnm{{Cattaneo}}},
\batitle{{Magnetic-Field Generation in Kolmogorov Turbulence}}.
\bjtitle{Physical Review Letters}
\bvolume{92}(\bissue{14}),
\bfpage{144501}
(\byear{2004})
\end{barticle}
\endbibitem

\bibitem[\protect\citeauthoryear{{Borrero} and {Kobel}}{2011}]{Borrero2011}
\begin{barticle}
\bauthor{\binits{J.M.} \bsnm{{Borrero}}},
\bauthor{\binits{P.} \bsnm{{Kobel}}},
\batitle{{Inferring the magnetic field vector in the quiet Sun. I. Photon noise
  and selection criteria}}.
\bjtitle{\aap}
\bvolume{527},
\bfpage{29}
(\byear{2011})
\end{barticle}
\endbibitem

\bibitem[\protect\citeauthoryear{{Brandenburg}}{2011}]{Branden2011}
\begin{barticle}
\bauthor{\binits{A.} \bsnm{{Brandenburg}}},
\batitle{{Nonlinear Small-scale Dynamos at Low Magnetic Prandtl Numbers}}.
\bjtitle{\apj}
\bvolume{741},
\bfpage{92}
(\byear{2011})
\end{barticle}
\endbibitem

\bibitem[\protect\citeauthoryear{{Cattaneo}}{1999}]{Cattaneo1999}
\begin{barticle}
\bauthor{\binits{F.} \bsnm{{Cattaneo}}},
\batitle{{On the Origin of Magnetic Fields in the Quiet Photosphere}}.
\bjtitle{\apjl}
\bvolume{515},
\bfpage{39}--\blpage{42}
(\byear{1999})
\end{barticle}
\endbibitem

\bibitem[\protect\citeauthoryear{{Centeno} et~al.}{2007}]{Rebeca2007}
\begin{barticle}
\bauthor{\binits{R.} \bsnm{{Centeno}}},
\bauthor{\binits{H.} \bsnm{{Socas-Navarro}}},
\bauthor{\binits{B.} \bsnm{{Lites}}},
\bauthor{\binits{M.} \bsnm{{Kubo}}},
\bauthor{\binits{Z.} \bsnm{{Frank}}},
\bauthor{\binits{R.} \bsnm{{Shine}}},
\bauthor{\binits{T.} \bsnm{{Tarbell}}},
\bauthor{\binits{A.} \bsnm{{Title}}},
\bauthor{\binits{K.} \bsnm{{Ichimoto}}},
\bauthor{\binits{S.} \bsnm{{Tsuneta}}},
\bauthor{\binits{Y.} \bsnm{{Katsukawa}}},
\bauthor{\binits{Y.} \bsnm{{Suematsu}}},
\bauthor{\binits{T.} \bsnm{{Shimizu}}},
\bauthor{\binits{S.} \bsnm{{Nagata}}},
\batitle{{Emergence of Small-Scale Magnetic Loops in the Quiet-Sun
  Internetwork}}.
\bjtitle{\apjl}
\bvolume{666},
\bfpage{137}--\blpage{140}
(\byear{2007})
\end{barticle}
\endbibitem

\bibitem[\protect\citeauthoryear{{Collados} et~al.}{2010}]{Collados2010}
\begin{barticle}
\bauthor{\binits{M.} \bsnm{{Collados}}},
\bauthor{\binits{F.} \bsnm{{Bettonvil}}},
\bauthor{\binits{L.} \bsnm{{Cavaller}}},
\bauthor{\binits{I.} \bsnm{{Ermolli}}},
\bauthor{\binits{B.} \bsnm{{Gelly}}},
\bauthor{\binits{A.} \bsnm{{P{\'e}rez}}},
\bauthor{\binits{H.} \bsnm{{Socas-Navarro}}},
\bauthor{\binits{D.} \bsnm{{Soltau}}},
\bauthor{\binits{R.} \bsnm{{Volkmer}}},
\bauthor{\bsnm{{EST Team}}},
\batitle{{European Solar Telescope: Progress status}}.
\bjtitle{Astronomische Nachrichten}
\bvolume{331},
\bfpage{615}
(\byear{2010})
\end{barticle}
\endbibitem

\bibitem[\protect\citeauthoryear{{Danilovic} et~al.}{2010a}]{Danilovic2010a}
\begin{barticle}
\bauthor{\binits{S.} \bsnm{{Danilovic}}},
\bauthor{\binits{M.} \bsnm{{Sch{\"u}ssler}}},
\bauthor{\binits{S.K.} \bsnm{{Solanki}}},
\batitle{{Probing quiet Sun magnetism using MURaM simulations and Hinode/SP
  results: support for a local dynamo}}.
\bjtitle{\aap}
\bvolume{513},
\bfpage{1}
(\byear{2010}a)
\end{barticle}
\endbibitem

\bibitem[\protect\citeauthoryear{{Danilovic} et~al.}{2010b}]{Danilovic2010b}
\begin{barticle}
\bauthor{\binits{S.} \bsnm{{Danilovic}}},
\bauthor{\binits{B.} \bsnm{{Beeck}}},
\bauthor{\binits{A.} \bsnm{{Pietarila}}},
\bauthor{\binits{M.} \bsnm{{Sch{\"u}ssler}}},
\bauthor{\binits{S.K.} \bsnm{{Solanki}}},
\bauthor{\binits{V.} \bsnm{{Mart{\'{\i}}nez Pillet}}},
\bauthor{\binits{J.A.} \bsnm{{Bonet}}},
\bauthor{\binits{J.C.} \bsnm{{del Toro Iniesta}}},
\bauthor{\binits{V.} \bsnm{{Domingo}}},
\bauthor{\binits{P.} \bsnm{{Barthol}}},
\bauthor{\binits{T.} \bsnm{{Berkefeld}}},
\bauthor{\binits{A.} \bsnm{{Gandorfer}}},
\bauthor{\binits{M.} \bsnm{{Kn{\"o}lker}}},
\bauthor{\binits{W.} \bsnm{{Schmidt}}},
\bauthor{\binits{A.M.} \bsnm{{Title}}},
\batitle{{Transverse Component of the Magnetic Field in the Solar Photosphere
  Observed by SUNRISE}}.
\bjtitle{\apjl}
\bvolume{723},
\bfpage{149}--\blpage{153}
(\byear{2010}b)
\end{barticle}
\endbibitem

\bibitem[\protect\citeauthoryear{{de Wijn} et~al.}{2009}]{dewijn2009}
\begin{barticle}
\bauthor{\binits{A.G.} \bsnm{{de Wijn}}},
\bauthor{\binits{J.O.} \bsnm{{Stenflo}}},
\bauthor{\binits{S.K.} \bsnm{{Solanki}}},
\bauthor{\binits{S.} \bsnm{{Tsuneta}}},
\batitle{{Small-Scale Solar Magnetic Fields}}.
\bjtitle{Space Science Reviews}
\bvolume{144},
\bfpage{275}--\blpage{315}
(\byear{2009})
\end{barticle}
\endbibitem

\bibitem[\protect\citeauthoryear{{Gandorfer} et~al.}{2011}]{Gandorfer2011}
\begin{barticle}
\bauthor{\binits{A.} \bsnm{{Gandorfer}}},
\bauthor{\binits{S.K.} \bsnm{{Solanki}}},
\bauthor{\binits{J.} \bsnm{{Woch}}},
\bauthor{\binits{V.} \bsnm{{Mart{\'{\i}}nez Pillet}}},
\bauthor{\binits{A.} \bsnm{{{\'A}lvarez Herrero}}},
\bauthor{\binits{T.} \bsnm{{Appourchaux}}},
\batitle{{The Solar Orbiter Mission and its Polarimetric and Helioseismic
  Imager (SO/PHI)}}.
\bjtitle{Journal of Physics Conference Series}
\bvolume{271}(\bissue{1}),
\bfpage{012086}
(\byear{2011})
\end{barticle}
\endbibitem

\bibitem[\protect\citeauthoryear{{G{\"o}m{\"o}ry} et~al.}{2010}]{Gomory2010}
\begin{barticle}
\bauthor{\binits{P.} \bsnm{{G{\"o}m{\"o}ry}}},
\bauthor{\binits{C.} \bsnm{{Beck}}},
\bauthor{\binits{H.} \bsnm{{Balthasar}}},
\bauthor{\binits{J.} \bsnm{{Ryb{\'a}k}}},
\bauthor{\binits{A.} \bsnm{{Ku{\v c}era}}},
\bauthor{\binits{J.} \bsnm{{Koza}}},
\bauthor{\binits{H.} \bsnm{{W{\"o}hl}}},
\batitle{{Magnetic loop emergence within a granule}}.
\bjtitle{\aap}
\bvolume{511},
\bfpage{14}
(\byear{2010})
\end{barticle}
\endbibitem

\bibitem[\protect\citeauthoryear{{Guglielmino} et~al.}{2012}]{Salvo2012}
\begin{barticle}
\bauthor{\binits{S.L.} \bsnm{{Guglielmino}}},
\bauthor{\binits{V.} \bsnm{{Mart{\'{\i}}nez Pillet}}},
\bauthor{\binits{J.A.} \bsnm{{Bonet}}},
\bauthor{\binits{J.C.} \bsnm{{del Toro Iniesta}}},
\bauthor{\binits{L.R.} \bsnm{{Bellot Rubio}}},
\bauthor{\binits{S.K.} \bsnm{{Solanki}}},
\bauthor{\binits{W.} \bsnm{{Schmidt}}},
\bauthor{\binits{A.} \bsnm{{Gandorfer}}},
\bauthor{\binits{P.} \bsnm{{Barthol}}},
\bauthor{\binits{M.} \bsnm{{Kn{\"o}lker}}},
\batitle{{The Frontier between Small-scale Bipoles and Ephemeral Regions in the
  Solar Photosphere: Emergence and Decay of an Intermediate-scale Bipole
  Observed with SUNRISE/IMaX}}.
\bjtitle{\apj}
\bvolume{745},
\bfpage{160}
(\byear{2012})
\end{barticle}
\endbibitem

\bibitem[\protect\citeauthoryear{{Hagenaar} et~al.}{2003}]{Hagenaar2003}
\begin{barticle}
\bauthor{\binits{H.J.} \bsnm{{Hagenaar}}},
\bauthor{\binits{C.J.} \bsnm{{Schrijver}}},
\bauthor{\binits{A.M.} \bsnm{{Title}}},
\batitle{{The Properties of Small Magnetic Regions on the Solar Surface and the
  Implications for the Solar Dynamo(s)}}.
\bjtitle{\apj}
\bvolume{584},
\bfpage{1107}--\blpage{1119}
(\byear{2003})
\end{barticle}
\endbibitem

\bibitem[\protect\citeauthoryear{{Harvey} et~al.}{2007}]{Harvey2007}
\begin{barticle}
\bauthor{\binits{J.W.} \bsnm{{Harvey}}},
\bauthor{\binits{D.} \bsnm{{Branston}}},
\bauthor{\binits{C.J.} \bsnm{{Henney}}},
\bauthor{\binits{C.U.} \bsnm{{Keller}}},
\bauthor{\bsnm{{SOLIS and GONG Teams}}},
\batitle{{Seething Horizontal Magnetic Fields in the Quiet Solar Photosphere}}.
\bjtitle{\apjl}
\bvolume{659},
\bfpage{177}--\blpage{180}
(\byear{2007})
\end{barticle}
\endbibitem

\bibitem[\protect\citeauthoryear{{Ishikawa} and {Tsuneta}}{2011}]{Ishikawa2011}
\begin{barticle}
\bauthor{\binits{R.} \bsnm{{Ishikawa}}},
\bauthor{\binits{S.} \bsnm{{Tsuneta}}},
\batitle{{The Relationship between Vertical and Horizontal Magnetic Fields in
  the Quiet Sun}}.
\bjtitle{\apj}
\bvolume{735},
\bfpage{74}
(\byear{2011})
\end{barticle}
\endbibitem

\bibitem[\protect\citeauthoryear{{Ishikawa} et~al.}{2010}]{Ishikawa2010}
\begin{barticle}
\bauthor{\binits{R.} \bsnm{{Ishikawa}}},
\bauthor{\binits{S.} \bsnm{{Tsuneta}}},
\bauthor{\binits{J.} \bsnm{{Jur{\v c}{\'a}k}}},
\batitle{{Three-Dimensional View of Transient Horizontal Magnetic Fields in the
  Photosphere}}.
\bjtitle{\apj}
\bvolume{713},
\bfpage{1310}--\blpage{1321}
(\byear{2010})
\end{barticle}
\endbibitem

\bibitem[\protect\citeauthoryear{{Iskakov} et~al.}{2007}]{Iska2007}
\begin{barticle}
\bauthor{\binits{A.B.} \bsnm{{Iskakov}}},
\bauthor{\binits{A.A.} \bsnm{{Schekochihin}}},
\bauthor{\binits{S.C.} \bsnm{{Cowley}}},
\bauthor{\binits{J.C.} \bsnm{{McWilliams}}},
\bauthor{\binits{M.R.E.} \bsnm{{Proctor}}},
\batitle{{Numerical Demonstration of Fluctuation Dynamo at Low Magnetic Prandtl
  Numbers}}.
\bjtitle{Physical Review Letters}
\bvolume{98}(\bissue{20}),
\bfpage{208501}
(\byear{2007})
\end{barticle}
\endbibitem

\bibitem[\protect\citeauthoryear{{Keil} et~al.}{2011}]{Keil2011}
\begin{bchapter}
\bauthor{\binits{S.L.} \bsnm{{Keil}}},
\bauthor{\binits{T.R.} \bsnm{{Rimmele}}},
\bauthor{\binits{J.} \bsnm{{Wagner}}},
\bauthor{\binits{D.} \bsnm{{Elmore}}},
\bauthor{\bsnm{{ATST Team}}},
\bctitle{{ATST: The Largest Polarimeter}},
in \bbtitle{Solar Polarization 6},
ed. by \beditor{\binits{J.R.} \bsnm{{Kuhn}}},
\beditor{\binits{D.M.} \bsnm{{Harrington}}},
\beditor{\binits{H.} \bsnm{{Lin}}},
\beditor{\binits{S.V.} \bsnm{{Berdyugina}}},
\beditor{\binits{J.} \bsnm{{Trujillo-Bueno}}},
\beditor{\binits{S.L.} \bsnm{{Keil}}},
\beditor{\binits{T.} \bsnm{{Rimmele}}}
\bsertitle{Astronomical Society of the Pacific Conference Series},
vol. \bseriesno{437},
\byear{2011},
p. \bfpage{319}
\end{bchapter}
\endbibitem

\bibitem[\protect\citeauthoryear{{Keller}}{1999}]{Keller1999}
\begin{bchapter}
\bauthor{\binits{C.} \bsnm{{Keller}}},
\bctitle{{The Advanced Solar Telescope: I. Science Goals}},
in \bbtitle{High Resolution Solar Physics: Theory, Observations, and
  Techniques},
ed. by \beditor{\binits{T.R.} \bsnm{{Rimmele}}},
\beditor{\binits{K.S.} \bsnm{{Balasubramaniam}}},
\beditor{\binits{R.R.} \bsnm{{Radick}}}
\bsertitle{Astronomical Society of the Pacific Conference Series},
vol. \bseriesno{183},
\byear{1999},
p. \bfpage{169}
\end{bchapter}
\endbibitem

\bibitem[\protect\citeauthoryear{{Keller} et~al.}{1994}]{Keller1994}
\begin{barticle}
\bauthor{\binits{C.U.} \bsnm{{Keller}}},
\bauthor{\binits{F.-L.} \bsnm{{Deubner}}},
\bauthor{\binits{U.} \bsnm{{Egger}}},
\bauthor{\binits{B.} \bsnm{{Fleck}}},
\bauthor{\binits{H.P.} \bsnm{{Povel}}},
\batitle{{On the strength of solar intra-network fields}}.
\bjtitle{\aap}
\bvolume{286},
\bfpage{626}--\blpage{634}
(\byear{1994})
\end{barticle}
\endbibitem

\bibitem[\protect\citeauthoryear{{Kleint} et~al.}{2011}]{Kleint2011}
\begin{barticle}
\bauthor{\binits{L.} \bsnm{{Kleint}}},
\bauthor{\binits{A.I.} \bsnm{{Shapiro}}},
\bauthor{\binits{S.V.} \bsnm{{Berdyugina}}},
\bauthor{\binits{M.} \bsnm{{Bianda}}},
\batitle{{Solar turbulent magnetic fields: Non-LTE modeling of the Hanle effect
  in the C$_{2}$ molecule}}.
\bjtitle{\aap}
\bvolume{536},
\bfpage{47}
(\byear{2011})
\end{barticle}
\endbibitem

\bibitem[\protect\citeauthoryear{{Kosugi} et~al.}{2007}]{Kosugi2007}
\begin{barticle}
\bauthor{\binits{T.} \bsnm{{Kosugi}}},
\bauthor{\binits{K.} \bsnm{{Matsuzaki}}},
\bauthor{\binits{T.} \bsnm{{Sakao}}},
\bauthor{\binits{T.} \bsnm{{Shimizu}}},
\bauthor{\binits{Y.} \bsnm{{Sone}}},
\bauthor{\binits{S.} \bsnm{{Tachikawa}}},
\bauthor{\binits{T.} \bsnm{{Hashimoto}}},
\bauthor{\binits{K.} \bsnm{{Minesugi}}},
\bauthor{\binits{A.} \bsnm{{Ohnishi}}},
\bauthor{\binits{T.} \bsnm{{Yamada}}},
\bauthor{\binits{S.} \bsnm{{Tsuneta}}},
\bauthor{\binits{H.} \bsnm{{Hara}}},
\bauthor{\binits{K.} \bsnm{{Ichimoto}}},
\bauthor{\binits{Y.} \bsnm{{Suematsu}}},
\bauthor{\binits{M.} \bsnm{{Shimojo}}},
\bauthor{\binits{T.} \bsnm{{Watanabe}}},
\bauthor{\binits{S.} \bsnm{{Shimada}}},
\bauthor{\binits{J.M.} \bsnm{{Davis}}},
\bauthor{\binits{L.D.} \bsnm{{Hill}}},
\bauthor{\binits{J.K.} \bsnm{{Owens}}},
\bauthor{\binits{A.M.} \bsnm{{Title}}},
\bauthor{\binits{J.L.} \bsnm{{Culhane}}},
\bauthor{\binits{L.K.} \bsnm{{Harra}}},
\bauthor{\binits{G.A.} \bsnm{{Doschek}}},
\bauthor{\binits{L.} \bsnm{{Golub}}},
\batitle{{The Hinode (Solar-B) Mission: An Overview}}.
\bjtitle{\solphys}
\bvolume{243},
\bfpage{3}--\blpage{17}
(\byear{2007})
\end{barticle}
\endbibitem

\bibitem[\protect\citeauthoryear{{Lin}}{1995}]{Lin1995}
\begin{barticle}
\bauthor{\binits{H.} \bsnm{{Lin}}},
\batitle{{On the Distribution of the Solar Magnetic Fields}}.
\bjtitle{\apj}
\bvolume{446},
\bfpage{421}
(\byear{1995})
\end{barticle}
\endbibitem

\bibitem[\protect\citeauthoryear{{Lites}}{2011}]{Lites2011}
\begin{barticle}
\bauthor{\binits{B.W.} \bsnm{{Lites}}},
\batitle{{Hinode Observations Suggesting the Presence of a Local Small-scale
  Turbulent Dynamo}}.
\bjtitle{\apj}
\bvolume{737},
\bfpage{52}
(\byear{2011})
\end{barticle}
\endbibitem

\bibitem[\protect\citeauthoryear{{Lites} et~al.}{2008}]{Lites2008}
\begin{barticle}
\bauthor{\binits{B.W.} \bsnm{{Lites}}}, \betal,
\batitle{{The Horizontal Magnetic Flux of the Quiet-Sun Internetwork as
  Observed with the Hinode Spectro-Polarimeter}}.
\bjtitle{\apj}
\bvolume{672},
\bfpage{1237}--\blpage{1253}
(\byear{2008})
\end{barticle}
\endbibitem

\bibitem[\protect\citeauthoryear{{Lites} et~al.}{1996}]{Lites1996}
\begin{barticle}
\bauthor{\binits{B.W.} \bsnm{{Lites}}},
\bauthor{\binits{K.D.} \bsnm{{Leka}}},
\bauthor{\binits{A.} \bsnm{{Skumanich}}},
\bauthor{\binits{V.} \bsnm{{Martinez Pillet}}},
\bauthor{\binits{T.} \bsnm{{Shimizu}}},
\batitle{{Small-Scale Horizontal Magnetic Fields in the Solar Photosphere}}.
\bjtitle{\apj}
\bvolume{460},
\bfpage{1019}
(\byear{1996})
\end{barticle}
\endbibitem

\bibitem[\protect\citeauthoryear{{Lites} et~al.}{2009}]{Lites2009}
\begin{bchapter}
\bauthor{\binits{B.W.} \bsnm{{Lites}}},
\bauthor{\binits{M.} \bsnm{{Kubo}}},
\bauthor{\binits{H.} \bsnm{{Socas-Navarro}}},
\bauthor{\binits{T.} \bsnm{{Berger}}},
\bauthor{\binits{Z.} \bsnm{{Frank}}},
\bauthor{\binits{R.} \bsnm{{Shine}}},
\bauthor{\binits{T.} \bsnm{{Tarbell}}},
\bauthor{\binits{A.M.} \bsnm{{Title}}},
\bauthor{\binits{K.} \bsnm{{Ichimoto}}},
\bauthor{\binits{Y.} \bsnm{{Katsukawa}}},
\bauthor{\binits{S.} \bsnm{{Tsuneta}}},
\bauthor{\binits{Y.} \bsnm{{Suematsu}}},
\bauthor{\binits{T.} \bsnm{{Shimizu}}},
\bauthor{\binits{S.} \bsnm{{Nagata}}},
\bctitle{{Has Hinode Revealed the Missing Turbulent Flux of the Quiet Sun?}},
in \bbtitle{Solar Polarization 5: In Honor of Jan Stenflo},
ed. by \beditor{\binits{S.V.} \bsnm{{Berdyugina}}},
\beditor{\binits{K.N.} \bsnm{{Nagendra}}},
\beditor{\binits{R.} \bsnm{{Ramelli}}}
\bsertitle{Astronomical Society of the Pacific Conference Series},
vol. \bseriesno{405},
\byear{2009},
p. \bfpage{173}
\end{bchapter}
\endbibitem

\bibitem[\protect\citeauthoryear{{Mart{\'{\i}}nez Gonz{\'a}lez} and {Bellot
  Rubio}}{2009}]{Marian2009}
\begin{barticle}
\bauthor{\binits{M.J.} \bsnm{{Mart{\'{\i}}nez Gonz{\'a}lez}}},
\bauthor{\binits{L.R.} \bsnm{{Bellot Rubio}}},
\batitle{{Emergence of Small-scale Magnetic Loops Through the Quiet Solar
  Atmosphere}}.
\bjtitle{\apj}
\bvolume{700},
\bfpage{1391}--\blpage{1403}
(\byear{2009})
\end{barticle}
\endbibitem

\bibitem[\protect\citeauthoryear{{Mart{\'{\i}}nez Gonz{\'a}lez}
  et~al.}{2012}]{Marian2012}
\begin{barticle}
\bauthor{\binits{M.J.} \bsnm{{Mart{\'{\i}}nez Gonz{\'a}lez}}},
\bauthor{\binits{R.} \bsnm{{Manso Sainz}}},
\bauthor{\binits{A.} \bsnm{{Asensio Ramos}}},
\bauthor{\binits{E.} \bsnm{{Hijano}}},
\batitle{{Dead Calm Areas in the Very Quiet Sun}}.
\bjtitle{\apj}
\bvolume{755},
\bfpage{175}
(\byear{2012})
\end{barticle}
\endbibitem

\bibitem[\protect\citeauthoryear{{Mart{\'{\i}}nez Pillet}
  et~al.}{2011}]{Pillet2011}
\begin{barticle}
\bauthor{\binits{V.} \bsnm{{Mart{\'{\i}}nez Pillet}}},
\bauthor{\binits{J.C.} \bsnm{{Del Toro Iniesta}}},
\bauthor{\binits{A.} \bsnm{{{\'A}lvarez-Herrero}}},
\bauthor{\binits{V.} \bsnm{{Domingo}}},
\bauthor{\binits{J.A.} \bsnm{{Bonet}}},
\bauthor{\binits{L.} \bsnm{{Gonz{\'a}lez Fern{\'a}ndez}}},
\bauthor{\binits{A.} \bsnm{{L{\'o}pez Jim{\'e}nez}}},
\bauthor{\binits{C.} \bsnm{{Pastor}}},
\bauthor{\binits{J.L.} \bsnm{{Gasent Blesa}}},
\bauthor{\binits{P.} \bsnm{{Mellado}}},
\bauthor{\binits{J.} \bsnm{{Piqueras}}},
\bauthor{\binits{B.} \bsnm{{Aparicio}}},
\bauthor{\binits{M.} \bsnm{{Balaguer}}},
\bauthor{\binits{E.} \bsnm{{Ballesteros}}},
\bauthor{\binits{T.} \bsnm{{Belenguer}}},
\bauthor{\binits{L.R.} \bsnm{{Bellot Rubio}}},
\bauthor{\binits{T.} \bsnm{{Berkefeld}}},
\bauthor{\binits{M.} \bsnm{{Collados}}},
\bauthor{\binits{W.} \bsnm{{Deutsch}}},
\bauthor{\binits{A.} \bsnm{{Feller}}},
\bauthor{\binits{F.} \bsnm{{Girela}}},
\bauthor{\binits{B.} \bsnm{{Grauf}}},
\bauthor{\binits{R.L.} \bsnm{{Heredero}}},
\bauthor{\binits{M.} \bsnm{{Herranz}}},
\bauthor{\binits{J.M.} \bsnm{{Jer{\'o}nimo}}},
\bauthor{\binits{H.} \bsnm{{Laguna}}},
\bauthor{\binits{R.} \bsnm{{Meller}}},
\bauthor{\binits{M.} \bsnm{{Men{\'e}ndez}}},
\bauthor{\binits{R.} \bsnm{{Morales}}},
\bauthor{\binits{D.} \bsnm{{Orozco Su{\'a}rez}}},
\bauthor{\binits{G.} \bsnm{{Ramos}}},
\bauthor{\binits{M.} \bsnm{{Reina}}},
\bauthor{\binits{J.L.} \bsnm{{Ramos}}},
\bauthor{\binits{P.} \bsnm{{Rodr{\'{\i}}guez}}},
\bauthor{\binits{A.} \bsnm{{S{\'a}nchez}}},
\bauthor{\binits{N.} \bsnm{{Uribe-Patarroyo}}},
\bauthor{\binits{P.} \bsnm{{Barthol}}},
\bauthor{\binits{A.} \bsnm{{Gandorfer}}},
\bauthor{\binits{M.} \bsnm{{Knoelker}}},
\bauthor{\binits{W.} \bsnm{{Schmidt}}},
\bauthor{\binits{S.K.} \bsnm{{Solanki}}},
\bauthor{\binits{S.} \bsnm{{Vargas Dom{\'{\i}}nguez}}},
\batitle{{The Imaging Magnetograph eXperiment (IMaX) for the Sunrise
  Balloon-Borne Solar Observatory}}.
\bjtitle{\solphys}
\bvolume{268},
\bfpage{57}--\blpage{102}
(\byear{2011})
\end{barticle}
\endbibitem

\bibitem[\protect\citeauthoryear{{Meunier} et~al.}{2007}]{Meunier2007}
\begin{barticle}
\bauthor{\binits{N.} \bsnm{{Meunier}}},
\bauthor{\binits{R.} \bsnm{{Tkaczuk}}},
\bauthor{\binits{T.} \bsnm{{Roudier}}},
\bauthor{\binits{M.} \bsnm{{Rieutord}}},
\batitle{{Velocities and divergences as a function of supergranule size}}.
\bjtitle{\aap}
\bvolume{461},
\bfpage{1141}--\blpage{1147}
(\byear{2007})
\end{barticle}
\endbibitem

\bibitem[\protect\citeauthoryear{{Moll} et~al.}{2011}]{Moll2011}
\begin{barticle}
\bauthor{\binits{R.} \bsnm{{Moll}}},
\bauthor{\binits{J.} \bsnm{{Pietarila Graham}}},
\bauthor{\binits{J.} \bsnm{{Pratt}}},
\bauthor{\binits{R.H.} \bsnm{{Cameron}}},
\bauthor{\binits{W.-C.} \bsnm{{M{\"u}ller}}},
\bauthor{\binits{M.} \bsnm{{Sch{\"u}ssler}}},
\batitle{{Universality of the Small-scale Dynamo Mechanism}}.
\bjtitle{\apj}
\bvolume{736},
\bfpage{36}
(\byear{2011})
\end{barticle}
\endbibitem

\bibitem[\protect\citeauthoryear{{M{\"u}ller} et~al.}{2012}]{Muller2012}
\begin{botherref}
\oauthor{\binits{D.} \bsnm{{M{\"u}ller}}},
\oauthor{\binits{R.G.} \bsnm{{Marsden}}},
\oauthor{\binits{O.C.} \bsnm{{St.~Cyr}}},
\oauthor{\binits{H.R.} \bsnm{{Gilbert}}},
{Solar Orbiter}.
\solphys,
193
(2012)
\end{botherref}
\endbibitem

\bibitem[\protect\citeauthoryear{{Nordlund} et~al.}{1992}]{Nordlund1992}
\begin{barticle}
\bauthor{\binits{A.} \bsnm{{Nordlund}}},
\bauthor{\binits{A.} \bsnm{{Brandenburg}}},
\bauthor{\binits{R.L.} \bsnm{{Jennings}}},
\bauthor{\binits{M.} \bsnm{{Rieutord}}},
\bauthor{\binits{J.} \bsnm{{Ruokolainen}}},
\bauthor{\binits{R.F.} \bsnm{{Stein}}},
\bauthor{\binits{I.} \bsnm{{Tuominen}}},
\batitle{{Dynamo action in stratified convection with overshoot}}.
\bjtitle{\apj}
\bvolume{392},
\bfpage{647}--\blpage{652}
(\byear{1992})
\end{barticle}
\endbibitem

\bibitem[\protect\citeauthoryear{{Orozco Su{\'a}rez} and {Bellot
  Rubio}}{2012}]{Orozco2012}
\begin{barticle}
\bauthor{\binits{D.} \bsnm{{Orozco Su{\'a}rez}}},
\bauthor{\binits{L.R.} \bsnm{{Bellot Rubio}}},
\batitle{{Analysis of Quiet-Sun Internetwork Magnetic Fields Based on Linear
  Polarization Signals}}.
\bjtitle{\apj}
\bvolume{751},
\bfpage{2}
(\byear{2012})
\end{barticle}
\endbibitem

\bibitem[\protect\citeauthoryear{{Orozco Su{\'a}rez} et~al.}{2007}]{Orozco2007}
\begin{barticle}
\bauthor{\binits{D.} \bsnm{{Orozco Su{\'a}rez}}},
\bauthor{\binits{L.R.} \bsnm{{Bellot Rubio}}},
\bauthor{\binits{J.C.} \bsnm{{del Toro Iniesta}}},
\bauthor{\binits{S.} \bsnm{{Tsuneta}}},
\bauthor{\binits{B.W.} \bsnm{{Lites}}},
\bauthor{\binits{K.} \bsnm{{Ichimoto}}},
\bauthor{\binits{Y.} \bsnm{{Katsukawa}}},
\bauthor{\binits{S.} \bsnm{{Nagata}}},
\bauthor{\binits{T.} \bsnm{{Shimizu}}},
\bauthor{\binits{R.A.} \bsnm{{Shine}}},
\bauthor{\binits{Y.} \bsnm{{Suematsu}}},
\bauthor{\binits{T.D.} \bsnm{{Tarbell}}},
\bauthor{\binits{A.M.} \bsnm{{Title}}},
\batitle{{Quiet-Sun Internetwork Magnetic Fields from the Inversion of Hinode
  Measurements}}.
\bjtitle{\apjl}
\bvolume{670},
\bfpage{61}--\blpage{64}
(\byear{2007})
\end{barticle}
\endbibitem

\bibitem[\protect\citeauthoryear{{Petrovay} and {Szakaly}}{1993}]{Petrovay1993}
\begin{barticle}
\bauthor{\binits{K.} \bsnm{{Petrovay}}},
\bauthor{\binits{G.} \bsnm{{Szakaly}}},
\batitle{{The origin of intranetwork fields: a small-scale solar dynamo}}.
\bjtitle{\aap}
\bvolume{274},
\bfpage{543}
(\byear{1993})
\end{barticle}
\endbibitem

\bibitem[\protect\citeauthoryear{{Pietarila Graham} et~al.}{2010}]{Pieta2010}
\begin{barticle}
\bauthor{\binits{J.} \bsnm{{Pietarila Graham}}},
\bauthor{\binits{R.} \bsnm{{Cameron}}},
\bauthor{\binits{M.} \bsnm{{Sch{\"u}ssler}}},
\batitle{{Turbulent Small-Scale Dynamo Action in Solar Surface Simulations}}.
\bjtitle{\apj}
\bvolume{714},
\bfpage{1606}--\blpage{1616}
(\byear{2010})
\end{barticle}
\endbibitem

\bibitem[\protect\citeauthoryear{{Pietarila Graham} et~al.}{2009a}]{Pieta2009b}
\begin{bchapter}
\bauthor{\binits{J.} \bsnm{{Pietarila Graham}}},
\bauthor{\binits{S.} \bsnm{{Danilovic}}},
\bauthor{\binits{M.} \bsnm{{Sch{\"u}ssler}}},
\bctitle{{The Small-scale Solar Surface Dynamo (keynote)}},
in \bbtitle{The Second Hinode Science Meeting: Beyond Discovery-Toward
  Understanding},
ed. by \beditor{\binits{B.} \bsnm{{Lites}}},
\beditor{\binits{M.} \bsnm{{Cheung}}},
\beditor{\binits{T.} \bsnm{{Magara}}},
\beditor{\binits{J.} \bsnm{{Mariska}}},
\beditor{\binits{K.} \bsnm{{Reeves}}}
\bsertitle{Astronomical Society of the Pacific Conference Series},
vol. \bseriesno{415},
\byear{2009}a,
p. \bfpage{43}
\end{bchapter}
\endbibitem

\bibitem[\protect\citeauthoryear{{Pietarila Graham} et~al.}{2009b}]{Pieta2009a}
\begin{barticle}
\bauthor{\binits{J.} \bsnm{{Pietarila Graham}}},
\bauthor{\binits{S.} \bsnm{{Danilovic}}},
\bauthor{\binits{M.} \bsnm{{Sch{\"u}ssler}}},
\batitle{{Turbulent Magnetic Fields in the Quiet Sun: Implications of Hinode
  Observations and Small-Scale Dynamo Simulations}}.
\bjtitle{\apj}
\bvolume{693},
\bfpage{1728}--\blpage{1735}
(\byear{2009}b)
\end{barticle}
\endbibitem

\bibitem[\protect\citeauthoryear{{Ruiz Cobo} and {del Toro
  Iniesta}}{1992}]{SIR1992}
\begin{barticle}
\bauthor{\binits{B.} \bsnm{{Ruiz Cobo}}},
\bauthor{\binits{J.C.} \bsnm{{del Toro Iniesta}}},
\batitle{{Inversion of Stokes profiles}}.
\bjtitle{\apj}
\bvolume{398},
\bfpage{375}--\blpage{385}
(\byear{1992})
\end{barticle}
\endbibitem

\bibitem[\protect\citeauthoryear{{S{\'a}nchez Almeida}}{2005}]{Almeida2005}
\begin{barticle}
\bauthor{\binits{J.} \bsnm{{S{\'a}nchez Almeida}}},
\batitle{{On the Sr I {$\lambda$}4607 {\AA} Hanle depolarization signals in the
  quiet Sun}}.
\bjtitle{\aap}
\bvolume{438},
\bfpage{727}--\blpage{732}
(\byear{2005})
\end{barticle}
\endbibitem

\bibitem[\protect\citeauthoryear{{S{\'a}nchez Almeida} and {Mart{\'{\i}}nez
  Gonz{\'a}lez}}{2011}]{Almeida2011}
\begin{bchapter}
\bauthor{\binits{J.} \bsnm{{S{\'a}nchez Almeida}}},
\bauthor{\binits{M.} \bsnm{{Mart{\'{\i}}nez Gonz{\'a}lez}}},
\bctitle{{The Magnetic Fields of the Quiet Sun}},
in \bbtitle{Solar Polarization 6},
ed. by \beditor{\binits{J.R.} \bsnm{{Kuhn}}},
\beditor{\binits{D.M.} \bsnm{{Harrington}}},
\beditor{\binits{H.} \bsnm{{Lin}}},
\beditor{\binits{S.V.} \bsnm{{Berdyugina}}},
\beditor{\binits{J.} \bsnm{{Trujillo-Bueno}}},
\beditor{\binits{S.L.} \bsnm{{Keil}}},
\beditor{\binits{T.} \bsnm{{Rimmele}}}
\bsertitle{Astronomical Society of the Pacific Conference Series},
vol. \bseriesno{437},
\byear{2011},
p. \bfpage{451}
\end{bchapter}
\endbibitem

\bibitem[\protect\citeauthoryear{{S{\'a}nchez Almeida}
  et~al.}{2003}]{Almeida2003}
\begin{barticle}
\bauthor{\binits{J.} \bsnm{{S{\'a}nchez Almeida}}},
\bauthor{\binits{T.} \bsnm{{Emonet}}},
\bauthor{\binits{F.} \bsnm{{Cattaneo}}},
\batitle{{Polarization of Photospheric Lines from Turbulent Dynamo
  Simulations}}.
\bjtitle{\apj}
\bvolume{585},
\bfpage{536}--\blpage{552}
(\byear{2003})
\end{barticle}
\endbibitem

\bibitem[\protect\citeauthoryear{{Schekochihin} et~al.}{2004a}]{Scheko2004b}
\begin{barticle}
\bauthor{\binits{A.A.} \bsnm{{Schekochihin}}},
\bauthor{\binits{S.C.} \bsnm{{Cowley}}},
\bauthor{\binits{J.L.} \bsnm{{Maron}}},
\bauthor{\binits{J.C.} \bsnm{{McWilliams}}},
\batitle{{Critical Magnetic Prandtl Number for Small-Scale Dynamo}}.
\bjtitle{Physical Review Letters}
\bvolume{92}(\bissue{5}),
\bfpage{054502}
(\byear{2004}a)
\end{barticle}
\endbibitem

\bibitem[\protect\citeauthoryear{{Schekochihin} et~al.}{2004b}]{Scheko2004a}
\begin{barticle}
\bauthor{\binits{A.A.} \bsnm{{Schekochihin}}},
\bauthor{\binits{S.C.} \bsnm{{Cowley}}},
\bauthor{\binits{S.F.} \bsnm{{Taylor}}},
\bauthor{\binits{J.L.} \bsnm{{Maron}}},
\bauthor{\binits{J.C.} \bsnm{{McWilliams}}},
\batitle{{Simulations of the Small-Scale Turbulent Dynamo}}.
\bjtitle{\apj}
\bvolume{612},
\bfpage{276}--\blpage{307}
(\byear{2004}b)
\end{barticle}
\endbibitem

\bibitem[\protect\citeauthoryear{{Schekochihin} et~al.}{2005}]{Scheko2005}
\begin{barticle}
\bauthor{\binits{A.A.} \bsnm{{Schekochihin}}},
\bauthor{\binits{N.E.L.} \bsnm{{Haugen}}},
\bauthor{\binits{A.} \bsnm{{Brandenburg}}},
\bauthor{\binits{S.C.} \bsnm{{Cowley}}},
\bauthor{\binits{J.L.} \bsnm{{Maron}}},
\bauthor{\binits{J.C.} \bsnm{{McWilliams}}},
\batitle{{The Onset of a Small-Scale Turbulent Dynamo at Low Magnetic Prandtl
  Numbers}}.
\bjtitle{\apjl}
\bvolume{625},
\bfpage{115}--\blpage{118}
(\byear{2005})
\end{barticle}
\endbibitem

\bibitem[\protect\citeauthoryear{{Schekochihin} et~al.}{2007}]{Scheko2007}
\begin{barticle}
\bauthor{\binits{A.A.} \bsnm{{Schekochihin}}},
\bauthor{\binits{A.B.} \bsnm{{Iskakov}}},
\bauthor{\binits{S.C.} \bsnm{{Cowley}}},
\bauthor{\binits{J.C.} \bsnm{{McWilliams}}},
\bauthor{\binits{M.R.E.} \bsnm{{Proctor}}},
\bauthor{\binits{T.A.} \bsnm{{Yousef}}},
\batitle{{Fluctuation dynamo and turbulent induction at low magnetic Prandtl
  numbers}}.
\bjtitle{New Journal of Physics}
\bvolume{9},
\bfpage{300}
(\byear{2007})
\end{barticle}
\endbibitem

\bibitem[\protect\citeauthoryear{{Scherrer} et~al.}{2012}]{Scherrer2012}
\begin{barticle}
\bauthor{\binits{P.H.} \bsnm{{Scherrer}}},
\bauthor{\binits{J.} \bsnm{{Schou}}},
\bauthor{\binits{R.I.} \bsnm{{Bush}}},
\bauthor{\binits{A.G.} \bsnm{{Kosovichev}}},
\bauthor{\binits{R.S.} \bsnm{{Bogart}}},
\bauthor{\binits{J.T.} \bsnm{{Hoeksema}}},
\bauthor{\binits{Y.} \bsnm{{Liu}}},
\bauthor{\binits{T.L.} \bsnm{{Duvall}}},
\bauthor{\binits{J.} \bsnm{{Zhao}}},
\bauthor{\binits{A.M.} \bsnm{{Title}}},
\bauthor{\binits{C.J.} \bsnm{{Schrijver}}},
\bauthor{\binits{T.D.} \bsnm{{Tarbell}}},
\bauthor{\binits{S.} \bsnm{{Tomczyk}}},
\batitle{{The Helioseismic and Magnetic Imager (HMI) Investigation for the
  Solar Dynamics Observatory (SDO)}}.
\bjtitle{\solphys}
\bvolume{275},
\bfpage{207}--\blpage{227}
(\byear{2012})
\end{barticle}
\endbibitem

\bibitem[\protect\citeauthoryear{{Sch{\"u}ssler} and
  {V{\"o}gler}}{2008}]{Schussler2008}
\begin{barticle}
\bauthor{\binits{M.} \bsnm{{Sch{\"u}ssler}}},
\bauthor{\binits{A.} \bsnm{{V{\"o}gler}}},
\batitle{{Strong horizontal photospheric magnetic field in a surface dynamo
  simulation}}.
\bjtitle{\aap}
\bvolume{481},
\bfpage{5}--\blpage{8}
(\byear{2008})
\end{barticle}
\endbibitem

\bibitem[\protect\citeauthoryear{{Shchukina} and {Trujillo
  Bueno}}{2011}]{Shchu2011}
\begin{barticle}
\bauthor{\binits{N.} \bsnm{{Shchukina}}},
\bauthor{\binits{J.} \bsnm{{Trujillo Bueno}}},
\batitle{{Determining the Magnetization of the Quiet Sun Photosphere from the
  Hanle Effect and Surface Dynamo Simulations}}.
\bjtitle{\apjl}
\bvolume{731},
\bfpage{21}
(\byear{2011})
\end{barticle}
\endbibitem

\bibitem[\protect\citeauthoryear{{Solanki}}{2009}]{Solanki2009}
\begin{bchapter}
\bauthor{\binits{S.K.} \bsnm{{Solanki}}},
\bctitle{{Photospheric Magnetic Field: Quiet Sun}},
in \bbtitle{Solar Polarization 5: In Honor of Jan Stenflo},
ed. by \beditor{\binits{S.V.} \bsnm{{Berdyugina}}},
\beditor{\binits{K.N.} \bsnm{{Nagendra}}},
\beditor{\binits{R.} \bsnm{{Ramelli}}}
\bsertitle{Astronomical Society of the Pacific Conference Series},
vol. \bseriesno{405},
\byear{2009},
p. \bfpage{135}
\end{bchapter}
\endbibitem

\bibitem[\protect\citeauthoryear{{Solanki} et~al.}{2010}]{Solanki2010}
\begin{barticle}
\bauthor{\binits{S.K.} \bsnm{{Solanki}}},
\bauthor{\binits{P.} \bsnm{{Barthol}}},
\bauthor{\binits{S.} \bsnm{{Danilovic}}},
\bauthor{\binits{A.} \bsnm{{Feller}}},
\bauthor{\binits{A.} \bsnm{{Gandorfer}}},
\bauthor{\binits{J.} \bsnm{{Hirzberger}}},
\bauthor{\binits{T.L.} \bsnm{{Riethm{\"u}ller}}},
\bauthor{\binits{M.} \bsnm{{Sch{\"u}ssler}}},
\bauthor{\binits{J.A.} \bsnm{{Bonet}}},
\bauthor{\binits{V.} \bsnm{{Mart{\'{\i}}nez Pillet}}},
\bauthor{\binits{J.C.} \bsnm{{del Toro Iniesta}}},
\bauthor{\binits{V.} \bsnm{{Domingo}}},
\bauthor{\binits{J.} \bsnm{{Palacios}}},
\bauthor{\binits{M.} \bsnm{{Kn{\"o}lker}}},
\bauthor{\binits{N.} \bsnm{{Bello Gonz{\'a}lez}}},
\bauthor{\binits{T.} \bsnm{{Berkefeld}}},
\bauthor{\binits{M.} \bsnm{{Franz}}},
\bauthor{\binits{W.} \bsnm{{Schmidt}}},
\bauthor{\binits{A.M.} \bsnm{{Title}}},
\batitle{{SUNRISE: Instrument, Mission, Data, and First Results}}.
\bjtitle{\apjl}
\bvolume{723},
\bfpage{127}--\blpage{133}
(\byear{2010})
\end{barticle}
\endbibitem

\bibitem[\protect\citeauthoryear{{Spruit}}{1974}]{Spruit1974}
\begin{barticle}
\bauthor{\binits{H.C.} \bsnm{{Spruit}}},
\batitle{{A model of the solar convection zone}}.
\bjtitle{\solphys}
\bvolume{34},
\bfpage{277}--\blpage{290}
(\byear{1974})
\end{barticle}
\endbibitem

\bibitem[\protect\citeauthoryear{{Stein}}{2012}]{Stein2012}
\begin{barticle}
\bauthor{\binits{R.F.} \bsnm{{Stein}}},
\batitle{{Solar Surface Magneto-Convection}}.
\bjtitle{Living Reviews in Solar Physics}
\bvolume{9},
\bfpage{4}
(\byear{2012})
\end{barticle}
\endbibitem

\bibitem[\protect\citeauthoryear{{Stein} and {Nordlund}}{2006}]{Stein2006}
\begin{barticle}
\bauthor{\binits{R.F.} \bsnm{{Stein}}},
\bauthor{\binits{{\AA}.} \bsnm{{Nordlund}}},
\batitle{{Solar Small-Scale Magnetoconvection}}.
\bjtitle{\apj}
\bvolume{642},
\bfpage{1246}--\blpage{1255}
(\byear{2006})
\end{barticle}
\endbibitem

\bibitem[\protect\citeauthoryear{{Stein} et~al.}{2003}]{Stein2003}
\begin{bchapter}
\bauthor{\binits{R.F.} \bsnm{{Stein}}},
\bauthor{\binits{D.} \bsnm{{Bercik}}},
\bauthor{\binits{{\AA}.} \bsnm{{Nordlund}}},
\bctitle{{Solar Surface Magneto-convection}},
in \bbtitle{Current Theoretical Models and Future High Resolution Solar
  Observations: Preparing for ATST},
ed. by \beditor{\binits{A.A.} \bsnm{{Pevtsov}}},
\beditor{\binits{H.} \bsnm{{Uitenbroek}}}
\bsertitle{Astronomical Society of the Pacific Conference Series},
vol. \bseriesno{286},
\byear{2003},
p. \bfpage{121}
\end{bchapter}
\endbibitem

\bibitem[\protect\citeauthoryear{{Steiner} and {Rezaei}}{2012}]{Steiner2012}
\begin{bchapter}
\bauthor{\binits{O.} \bsnm{{Steiner}}},
\bauthor{\binits{R.} \bsnm{{Rezaei}}},
\bctitle{{Recent Advances in the Exploration of the Small-scale Structure of
  the Quiet Solar Atmosphere: Vortex Flows, the Horizontal Magnetic Field, and
  the Stokes- V Line-ratio Method}},
in \bbtitle{Fifth Hinode Science Meeting},
ed. by \beditor{\binits{L.} \bsnm{{Golub}}},
\beditor{\binits{I.} \bsnm{{De Moortel}}},
\beditor{\binits{T.} \bsnm{{Shimizu}}}
\bsertitle{Astronomical Society of the Pacific Conference Series},
vol. \bseriesno{456},
\byear{2012},
p. \bfpage{3}
\end{bchapter}
\endbibitem

\bibitem[\protect\citeauthoryear{{Steiner} et~al.}{2008}]{Steiner2008}
\begin{barticle}
\bauthor{\binits{O.} \bsnm{{Steiner}}},
\bauthor{\binits{R.} \bsnm{{Rezaei}}},
\bauthor{\binits{W.} \bsnm{{Schaffenberger}}},
\bauthor{\binits{S.} \bsnm{{Wedemeyer-B{\"o}hm}}},
\batitle{{The Horizontal Internetwork Magnetic Field: Numerical Simulations in
  Comparison to Observations with Hinode}}.
\bjtitle{\apjl}
\bvolume{680},
\bfpage{85}--\blpage{88}
(\byear{2008})
\end{barticle}
\endbibitem

\bibitem[\protect\citeauthoryear{{Stenflo}}{1982}]{Stenflo1982}
\begin{barticle}
\bauthor{\binits{J.O.} \bsnm{{Stenflo}}},
\batitle{{The Hanle effect and the diagnostics of turbulent magnetic fields in
  the solar atmosphere}}.
\bjtitle{\solphys}
\bvolume{80},
\bfpage{209}--\blpage{226}
(\byear{1982})
\end{barticle}
\endbibitem

\bibitem[\protect\citeauthoryear{{Stenflo}}{2011}]{Stenflo2011}
\begin{bchapter}
\bauthor{\binits{J.O.} \bsnm{{Stenflo}}},
\bctitle{{Unsolved Problems in Solar Polarization}},
in \bbtitle{Solar Polarization 6},
ed. by \beditor{\binits{J.R.} \bsnm{{Kuhn}}},
\beditor{\binits{D.M.} \bsnm{{Harrington}}},
\beditor{\binits{H.} \bsnm{{Lin}}},
\beditor{\binits{S.V.} \bsnm{{Berdyugina}}},
\beditor{\binits{J.} \bsnm{{Trujillo-Bueno}}},
\beditor{\binits{S.L.} \bsnm{{Keil}}},
\beditor{\binits{T.} \bsnm{{Rimmele}}}
\bsertitle{Astronomical Society of the Pacific Conference Series},
vol. \bseriesno{437},
\byear{2011},
p. \bfpage{3}
\end{bchapter}
\endbibitem

\bibitem[\protect\citeauthoryear{{Stenflo}}{2012}]{Stenflo2012}
\begin{botherref}
\oauthor{\binits{J.O.} \bsnm{{Stenflo}}},
{Basal magnetic flux and the local solar dynamo}.
ArXiv e-prints
(2012)
\end{botherref}
\endbibitem

\bibitem[\protect\citeauthoryear{{Title} et~al.}{1989}]{Title1989}
\begin{barticle}
\bauthor{\binits{A.M.} \bsnm{{Title}}},
\bauthor{\binits{T.D.} \bsnm{{Tarbell}}},
\bauthor{\binits{K.P.} \bsnm{{Topka}}},
\bauthor{\binits{S.H.} \bsnm{{Ferguson}}},
\bauthor{\binits{R.A.} \bsnm{{Shine}}},
\bauthor{\bsnm{{SOUP Team}}},
\batitle{{Statistical properties of solar granulation derived from the SOUP
  instrument on Spacelab 2}}.
\bjtitle{\apj}
\bvolume{336},
\bfpage{475}--\blpage{494}
(\byear{1989})
\end{barticle}
\endbibitem

\bibitem[\protect\citeauthoryear{{Trujillo Bueno}}{2011}]{Trujillo2011}
\begin{bchapter}
\bauthor{\binits{J.} \bsnm{{Trujillo Bueno}}},
\bctitle{{Modeling Scattering Polarization for Probing Solar Magnetism}},
in \bbtitle{Solar Polarization 6},
ed. by \beditor{\binits{J.R.} \bsnm{{Kuhn}}},
\beditor{\binits{D.M.} \bsnm{{Harrington}}},
\beditor{\binits{H.} \bsnm{{Lin}}},
\beditor{\binits{S.V.} \bsnm{{Berdyugina}}},
\beditor{\binits{J.} \bsnm{{Trujillo-Bueno}}},
\beditor{\binits{S.L.} \bsnm{{Keil}}},
\beditor{\binits{T.} \bsnm{{Rimmele}}}
\bsertitle{Astronomical Society of the Pacific Conference Series},
vol. \bseriesno{437},
\byear{2011},
p. \bfpage{83}
\end{bchapter}
\endbibitem

\bibitem[\protect\citeauthoryear{{Trujillo Bueno} et~al.}{2006}]{Trujillo2006}
\begin{bchapter}
\bauthor{\binits{J.} \bsnm{{Trujillo Bueno}}},
\bauthor{\binits{A.} \bsnm{{Asensio Ramos}}},
\bauthor{\binits{N.} \bsnm{{Shchukina}}},
\bctitle{{The Hanle Effect in Atomic and Molecular Lines: A New Look at the
  Sun's Hidden Magnetism}},
in \bbtitle{Astronomical Society of the Pacific Conference Series},
ed. by \beditor{\binits{R.} \bsnm{{Casini}}},
\beditor{\binits{B.W.} \bsnm{{Lites}}}
\bsertitle{Astronomical Society of the Pacific Conference Series},
vol. \bseriesno{358},
\byear{2006},
p. \bfpage{269}
\end{bchapter}
\endbibitem

\bibitem[\protect\citeauthoryear{{Trujillo Bueno} et~al.}{2004}]{Trujillo2004}
\begin{barticle}
\bauthor{\binits{J.} \bsnm{{Trujillo Bueno}}},
\bauthor{\binits{N.} \bsnm{{Shchukina}}},
\bauthor{\binits{A.} \bsnm{{Asensio Ramos}}},
\batitle{{A substantial amount of hidden magnetic energy in the quiet Sun}}.
\bjtitle{Nature}
\bvolume{430},
\bfpage{326}--\blpage{329}
(\byear{2004})
\end{barticle}
\endbibitem

\bibitem[\protect\citeauthoryear{{Tsuneta} et~al.}{2008}]{Tsuneta2008}
\begin{barticle}
\bauthor{\binits{S.} \bsnm{{Tsuneta}}},
\bauthor{\binits{K.} \bsnm{{Ichimoto}}},
\bauthor{\binits{Y.} \bsnm{{Katsukawa}}},
\bauthor{\binits{S.} \bsnm{{Nagata}}},
\bauthor{\binits{M.} \bsnm{{Otsubo}}},
\bauthor{\binits{T.} \bsnm{{Shimizu}}},
\bauthor{\binits{Y.} \bsnm{{Suematsu}}},
\bauthor{\binits{M.} \bsnm{{Nakagiri}}},
\bauthor{\binits{M.} \bsnm{{Noguchi}}},
\bauthor{\binits{T.} \bsnm{{Tarbell}}},
\bauthor{\binits{A.} \bsnm{{Title}}},
\bauthor{\binits{R.} \bsnm{{Shine}}},
\bauthor{\binits{W.} \bsnm{{Rosenberg}}},
\bauthor{\binits{C.} \bsnm{{Hoffmann}}},
\bauthor{\binits{B.} \bsnm{{Jurcevich}}},
\bauthor{\binits{G.} \bsnm{{Kushner}}},
\bauthor{\binits{M.} \bsnm{{Levay}}},
\bauthor{\binits{B.} \bsnm{{Lites}}},
\bauthor{\binits{D.} \bsnm{{Elmore}}},
\bauthor{\binits{T.} \bsnm{{Matsushita}}},
\bauthor{\binits{N.} \bsnm{{Kawaguchi}}},
\bauthor{\binits{H.} \bsnm{{Saito}}},
\bauthor{\binits{I.} \bsnm{{Mikami}}},
\bauthor{\binits{L.D.} \bsnm{{Hill}}},
\bauthor{\binits{J.K.} \bsnm{{Owens}}},
\batitle{{The Solar Optical Telescope for the Hinode Mission: An Overview}}.
\bjtitle{\solphys}
\bvolume{249},
\bfpage{167}--\blpage{196}
(\byear{2008})
\end{barticle}
\endbibitem

\bibitem[\protect\citeauthoryear{{Viticchi{\'e}}}{2012}]{Viticchie2012}
\begin{barticle}
\bauthor{\binits{B.} \bsnm{{Viticchi{\'e}}}},
\batitle{{On the Polarimetric Signature of Emerging Magnetic Loops in the Quiet
  Sun}}.
\bjtitle{\apjl}
\bvolume{747},
\bfpage{36}
(\byear{2012})
\end{barticle}
\endbibitem

\bibitem[\protect\citeauthoryear{{V{\"o}gler} and
  {Sch{\"u}ssler}}{2007}]{Vogler2007}
\begin{barticle}
\bauthor{\binits{A.} \bsnm{{V{\"o}gler}}},
\bauthor{\binits{M.} \bsnm{{Sch{\"u}ssler}}},
\batitle{{A solar surface dynamo}}.
\bjtitle{\aap}
\bvolume{465},
\bfpage{43}--\blpage{46}
(\byear{2007})
\end{barticle}
\endbibitem

\bibitem[\protect\citeauthoryear{{Westendorp Plaza}
  et~al.}{1998}]{Westendorp1998}
\begin{barticle}
\bauthor{\binits{C.} \bsnm{{Westendorp Plaza}}},
\bauthor{\binits{J.C.} \bsnm{{del Toro Iniesta}}},
\bauthor{\binits{B.} \bsnm{{Ruiz Cobo}}},
\bauthor{\binits{V.} \bsnm{{Martinez Pillet}}},
\bauthor{\binits{B.W.} \bsnm{{Lites}}},
\bauthor{\binits{A.} \bsnm{{Skumanich}}},
\batitle{{Optical Tomography of a Sunspot. I. Comparison between Two Inversion
  Techniques}}.
\bjtitle{\apj}
\bvolume{494},
\bfpage{453}
(\byear{1998})
\end{barticle}
\endbibitem

\bibitem[\protect\citeauthoryear{{Yelles Chaouche} et~al.}{2011}]{Lotfi2011}
\begin{barticle}
\bauthor{\binits{L.} \bsnm{{Yelles Chaouche}}},
\bauthor{\binits{F.} \bsnm{{Moreno-Insertis}}},
\bauthor{\binits{V.} \bsnm{{Mart{\'{\i}}nez Pillet}}},
\bauthor{\binits{T.} \bsnm{{Wiegelmann}}},
\bauthor{\binits{J.A.} \bsnm{{Bonet}}},
\bauthor{\binits{M.} \bsnm{{Kn{\"o}lker}}},
\bauthor{\binits{L.R.} \bsnm{{Bellot Rubio}}},
\bauthor{\binits{J.C.} \bsnm{{del Toro Iniesta}}},
\bauthor{\binits{P.} \bsnm{{Barthol}}},
\bauthor{\binits{A.} \bsnm{{Gandorfer}}},
\bauthor{\binits{W.} \bsnm{{Schmidt}}},
\bauthor{\binits{S.K.} \bsnm{{Solanki}}},
\batitle{{Mesogranulation and the Solar Surface Magnetic Field Distribution}}.
\bjtitle{\apj}
\bvolume{727},
\bfpage{30}
(\byear{2011})
\end{barticle}
\endbibitem

\end{thebibliography}

\end{document}